\RequirePackage{silence}
\documentclass[aps,prl,onecolumn,10pt,amsmath,amssymb,superscriptaddress,nofootinbib,floatfix]{revtex4-2}

\usepackage{bm}
\usepackage{graphicx}
\usepackage[dvipsnames]{xcolor}
\usepackage{mathtools}
\usepackage{amsthm}
\usepackage{enumitem}
\usepackage{dsfont}
\usepackage[margin=1in]{geometry}   
\usepackage[caption=false]{subfig}
\usepackage{subcaption}
\usepackage[normalem]{ulem}
\usepackage{ragged2e}
\usepackage[justification=Justified, singlelinecheck=false]{caption}
\usepackage[nopatch=footnote]{microtype}
\usepackage{cancel}
\usepackage[hidelinks]{hyperref} 

\newcommand{\CV}{\mathrm{CV}}    
\newtheorem{theorem}{Theorem}[section]
\newtheorem{proposition}[theorem]{Proposition}
\newtheorem{lemma}[theorem]{Lemma}
\newtheorem{corollary}[theorem]{Corollary}
\newtheorem{remark}[theorem]{Remark}

\begin{document}
	\title{The impact of DNA replication on the timing of cellular events}	
	\author{Juan David Marmolejo-Lozano}
	\affiliation{School of Biological Sciences, The University of Edinburgh, Edinburgh, United Kingdom}
	\author{Nikola Popović}
	\affiliation{School of Mathematics, The University of Edinburgh, Edinburgh, United Kingdom}
	\affiliation{Maxwell Institute for Mathematical Sciences, The University of Edinburgh, Edinburgh, United Kingdom}
    \author{Ramon Grima}
    \email{ramon.grima@stonybrook.edu}
    \altaffiliation[present address: ]{Laufer Center for Physical and Quantitative Biology, Department of Applied Mathematics and Statistics, Stony Brook University, New York, United States of America}
    \affiliation{School of Biological Sciences, The University of Edinburgh, Edinburgh, United Kingdom}
    
	\begin{abstract}
		Cellular timing often relies on the accumulation of a regulatory protein to a threshold. For a stable protein produced at a constant rate, the threshold alone sets timing precision: the squared coefficient of variation of the first-passage time is the inverse of the threshold number of molecules. This counting limit appears to be a fundamental constraint. However, production rates vary across the cell cycle. By solving exactly a stochastic model coupling DNA replication and protein accumulation, we show that when a gene replicates at an optimal time, precision exceeds the counting limit.
	\end{abstract}
	
	\maketitle
	
Gene expression is intrinsically stochastic: the individual biochemical events underlying it, such as transcription, translation, and degradation, occur at random times. This randomness in event timing gives rise to cell-to-cell variability in molecule numbers \cite{Swain2002, Elowitz2002}. Correspondingly, the time required to reach a given molecule number threshold is a random variable that is governed by first-passage-time (FPT) statistics \cite{Ghusinga2017, Rijal2022, Ham2024}. A common strategy by which cells schedule events is to accumulate a regulatory protein until its copy number first reaches a threshold, triggering a downstream process \cite{Alon2019}. Accurate timing is often essential to biological function. For example, during embryonic development, \textit{Hox} genes are switched on in a strict temporal sequence~\cite{Lappin2006-ao,Montavon2014}, with premature activation of some of them resulting in a truncated body axis~\cite{Young2009}. Timing is equally critical in the cell cycle, where each phase transition is triggered when phase-specific cyclins accumulate beyond a threshold level~\cite{Matthews2022}. In both cases, it is not the average amount of a gene product that is decisive but, rather, the timing of the event and its reproducibility across cells. The key quantity is therefore the time at which product first reaches a triggering threshold, and its variability across the cell population.
    
The simplest reference case for this scenario is a stable protein produced by a single Poisson process at {\it{constant rate}} $k$. The time it takes to accumulate $\lambda$ molecules is then a sum of $\lambda$ independent exponential waiting times, i.e., Erlang distributed, with mean $\lambda/k$ and squared coefficient of variation $\CV^2=1/\lambda$ \cite{Ghusinga2017}, which we henceforth refer to as the \emph{counting limit}. 

However, this picture is incomplete, as protein production rates are not constant across the cell cycle: when a gene is replicated at a well-defined point in the cycle, its copy number doubles and the effective protein production rate also increases. This increase is roughly twofold in some cases~\cite{Zopf2013}; however, for many genes, it is significantly reduced by dosage compensation~\cite{Voichek2016}. How replication affects copy-number and concentration distributions has been studied using a variety of stochastic models ~\cite{Cao2020,Beentjes2020,JiaSinghGrima2022,Jia2023,peterson2015effects,dessalles2020models}. By contrast, the effect of replication on the \emph{timing} of threshold crossing has been examined only recently, in a model where replication is triggered when the gene’s own mRNA count reaches a fixed value and dosage is not compensated~\cite{Verma2026}. In reality, however, a gene’s replication time is set by the cell-cycle programme, independently of its own expression. 

Here, we study a general model in which replication is timed independently of copy-number fluctuations and gene dosage may be partially compensated. We obtain the exact FPT distribution for the protein copy number to reach a threshold, and we crucially show that the coupling of protein accumulation to DNA replication can push the noise in the FPT \emph{below} the counting limit $1/\lambda$. Furthermore, there is an optimal mean replication time at which the timing precision is maximal; when replication is itself the noisier clock, the coupling can instead amplify the noise above that limit.
    
\begin{figure}[t]
	\centering
	\includegraphics[width=0.8\columnwidth]{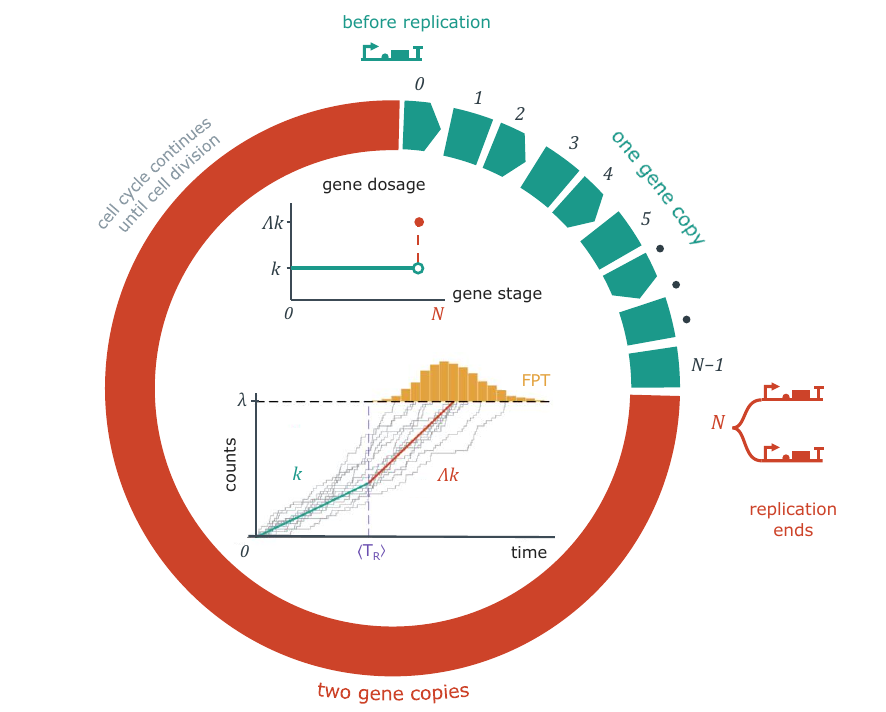}
	\caption{\textbf{Illustration of the model coupling protein accumulation to DNA replication}. Outer ring: with one gene copy (teal), replication advances through $N$ exponential steps at rate $\alpha = N/\langle T_R\rangle$, where $\langle T_R \rangle$ is the mean time from $t=0$ until the gene is replicated; replication of the gene ends when step $N$ is reached, and the cell cycle continues with two gene copies (red) until division. Insets: gene dosage (top), where replication raises the production rate from $k$ to $\Lambda k$, with $\Lambda \ge 1$; protein counts (bottom) accumulate at these rates up to the threshold $\lambda$. Grey: single trajectories; solid lines: mean counts before and after replication; orange: distribution of the FPT $T$ at $\lambda$.}
	\label{fig:schematic}
\end{figure}

\emph{Model}---We model a gene $G$ that synthesises a stable protein $P$ and undergoes replication as an $N$-step process (Fig.~\ref{fig:schematic}):
\begin{align}
   \label{eq:scheme}
   &G_0 \xrightarrow{\alpha} G_1 \xrightarrow{\alpha} \cdots \xrightarrow{\alpha} G_N,
   \quad
   G_{j<N} \xrightarrow{k} G_{j<N}+P, \\
   &G_N \xrightarrow{\Lambda k} G_N+P. \nonumber
\end{align}
The gene advances through the states $G_0\to G_1\to\cdots\to G_N$ at some constant rate $\alpha$; the corresponding $N$ steps represent progression from the initial time $t=0$ to replication of the gene, which doubles its copy number in the final transition ($N-1 \to N$). The time to DNA replication is therefore Erlang distributed, $T_R \sim \mathrm{Erlang}(N, N/\langle T_R \rangle)$, with mean $\langle T_R\rangle=N/\alpha$ and squared coefficient of variation $\CV_{T_R}^2=1/N$. Hence, $\langle T_R \rangle$ is the mean time from the onset of the replication process at $t=0$ until the gene is replicated. The multistep, constant-rate setup in Eq.~\eqref{eq:scheme} reproduces the non-exponential, Erlang-like distributions of cell-cycle and phase durations measured experimentally \cite{Yates2017,Chao2019}, and has also been applied in models of gene expression with replication and dosage compensation~\cite{Beentjes2020,Cao2020,JiaSinghGrima2022,sun2019nonlinear}. In these models, replication occurs after a fixed number of steps, i.e., at a reproducible fraction of the cell cycle, as in the stretched cell cycle model used to describe eukaryotic cells~\cite{Jia2023, Dowling2014}; correspondingly, we take $N$ steps to represent that fraction of the cell-cycle progression. Protein $P$ is produced at rate $k$ before replication and at the amplified rate $\Lambda k$ after. The dosage factor $\Lambda$ captures the change in gene dosage (Fig.~\ref{fig:schematic}, top inset): $\Lambda=2$ corresponds to exact copy-number doubling, while dosage-compensation mechanisms that partially buffer the extra gene copy place $\Lambda$ between $1$ and $2$ \cite{Voichek2016,Skinner2016, Kalita2025, Sukys2025}; our analysis holds for any $\Lambda \geq 1$. We assume protein $P$ to be stable on the timescale of interest. Since over 70\% of proteins in mouse fibroblasts have half-lives exceeding the cell cycle duration of around 27 hours~\cite{Schwanhusser2011}, degradation is negligible when the threshold must be reached well within a cycle. Similarly, in growing \textit{E. coli}, most proteins are not actively degraded \cite{Mosteller1980, Gupta2024} within a single cell cycle. We assume that at $t=0$ -- i.e., at the onset of both protein accumulation and the replication clock -- the gene is in state $G_0$ and that no copies of $P$ are present. We then solve for the distribution of the FPT $T$ at which the copy number of $P$ first reaches $\lambda$. 

\emph{Exact first-passage-time distribution}---The joint probability of the protein count and the replication state obeys a chemical master equation \cite{vanKampen2007}, which we solve exactly using the Laplace transform;  see the Supplemental Material for the derivation. Denoting by $\tilde h(s)=\int_0^\infty e^{-st}h(t)\,dt$ the transform of the FPT density $h(t)$, we obtain
\begin{align}
   &\tilde h(s) = \left(\frac{k}{s+k}\right)^{\!\lambda} I_p(\lambda,N) + 
   \nonumber\\
   &\left(\frac{k\Lambda}{s+k\Lambda}\right)^{\!\lambda}
   \left(\frac{N\Lambda/\langle T_R\rangle}
   {s(\Lambda-1)+N\Lambda/\langle T_R\rangle}\right)^{\!N}
   I_{p'}(N,\lambda),
   \label{eq:fpt}
\end{align}  
where
\begin{equation}
   p=\frac{s+k}{s+k+N/\langle T_R\rangle},
   \quad
   p'=\frac{s(\Lambda-1)+N\Lambda/\langle T_R\rangle}
   {\left[s+k+N/\langle T_R\rangle\right]\Lambda},
   \label{eq:pp}
\end{equation}
and $I_x(a,b)$ is the regularised incomplete Beta function \cite{NIST:DLMF}. The two terms in Eq.~\eqref{eq:fpt} admit a direct physical interpretation (Fig.~\ref{fig:schematic}, bottom inset). The first term describes threshold crossing \emph{before} replication, with protein produced at rate $k$: $[k/(s+k)]^{\lambda}$ is the Laplace transform of $\mathrm{Erlang}(\lambda,k)$, while $I_{p}(\lambda,N)$ is the probability, at $s=0$, that $\lambda$ molecules are produced before $N$ replication steps are completed. The second term describes crossing \emph{after} replication, and is a convolution of two clocks: an effective replication delay, $\mathrm{Erlang}$ of shape $N$ with mean $\mu_3=\langle T_R\rangle(\Lambda-1)/\Lambda$ and squared coefficient of variation $1/N$, and the subsequent accumulation at the boosted rate $k\Lambda$, $\mathrm{Erlang}$ of shape $\lambda$, with squared coefficient of variation $1/\lambda$. Correspondingly, $\mu_3$ represents the mean time ``lost" due to protein production at rate $k$, rather than $\Lambda k$, until replication. The FPT thus reflects the competition between these two clocks; which clock dominates, and how noisy each is, determines whether coupling decreases or increases timing precision.
	
\emph{Moments}---From $\langle T^m\rangle=(-1)^m\,\tilde h^{(m)}(0)$, we can obtain all moments of the FPT distribution in closed form; see the Supplemental Material. The mean is a weighted sum,
\begin{equation}
	\langle T\rangle=
	\mu_1\, I_{p_0}(\lambda,N)
	+(\mu_2+\mu_3)\big[1-I_{p_0}(\lambda,N)\big]
	+C_1\phi,
	\label{eq:mean}
\end{equation}
where $\mu_1=\lambda/k$ and $\mu_2=\lambda/(k\Lambda)$ are the Erlang means at low and high production rates, respectively; $\mu_3$ is the effective delay defined above; $p_0=k\langle T_R\rangle/(k\langle T_R\rangle+N) = p(s=0)$; and $C_1\phi$ corrects for the dependence between the route taken -- accumulation or replication -- and its duration. 
Finally, $I_{p_0}$ and $1-I_{p_0}$ are the probabilities that the threshold is reached before and after replication, respectively.

The same two-route structure is found for the second moment,
\begin{align}
	\label{eq:second}
	\langle T^2\rangle &=
	\langle t^2\rangle_{(1)}\, I_{p_0}(\lambda,N)
	\\ \nonumber
	&\quad+\big(\langle t^2\rangle_{(2)}+2\mu_2\mu_3+\langle t^2\rangle_{(3)}\big)
	\bigl[1-I_{p_0}(\lambda,N)\bigr]\\
	&\quad+C_2\phi, \nonumber
\end{align}
with $\langle t^2\rangle_{(l)}=\mu_l^2\,(1+\CV_l^2)$ ($l=1,2,3$), where $\CV_1^2=\CV_2^2=1/\lambda$ and $\CV_3^2=1/N$ are the squared coefficients of variation of the two accumulation clocks and of the effective replication delay, respectively; the cross-term $2\mu_2\mu_3$ derives from the convolution of the replication delay with the post-replication run to threshold. Here, $\phi=p_0^{\lambda-1}(1-p_0)^{N-1}/\mathcal{B}(\lambda,N)$ is the associated Beta density, while $C_1$ and $C_2$ are rational functions of the model parameters, given in the Supplemental Material. Finally, the correction $C_2\phi$ plays the role of $C_1\phi$ in Eq.~\eqref{eq:mean}.

The size of the timing noise then follows in closed form as $\CV_T^2=\langle T^2\rangle/\langle T\rangle^2-1$. In the two extremes, where replication is instantaneous ($\langle T_R\rangle\to0^+$, with $T\sim\mathrm{Erlang}(\lambda,k\Lambda)$) or where it is irrelevant ($\langle T_R\rangle\to\infty$, with $T\sim\mathrm{Erlang}(\lambda,k)$), we recover the counting limit: $\CV_T^2=1/\lambda$. Hence, any departure from that limit must occur at intermediate replication times.
    
\begin{figure*}[!t]
        \centering
        \subfloat[]{\includegraphics[width=\textwidth]{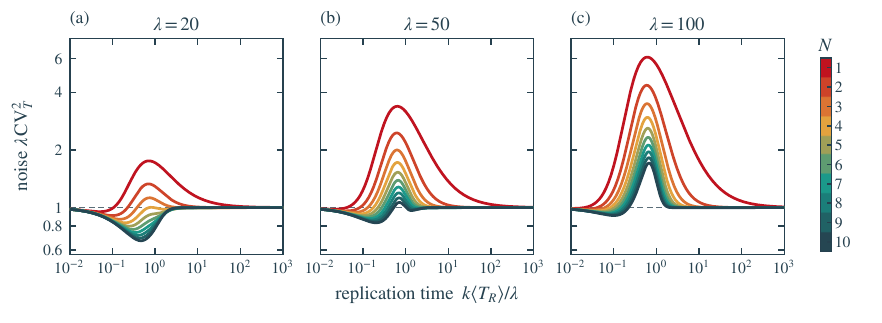}
        \label{fig:cv2}}\\[0.3em]
        \subfloat[]{\includegraphics[width=0.32\textwidth]{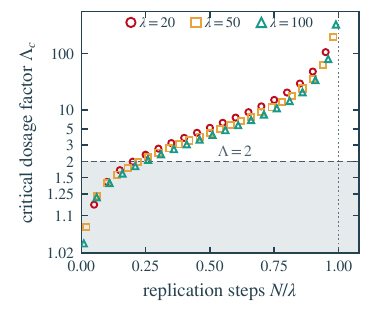}
        \label{fig:phase}}\hfil
        \subfloat[]{\includegraphics[width=0.32\textwidth]{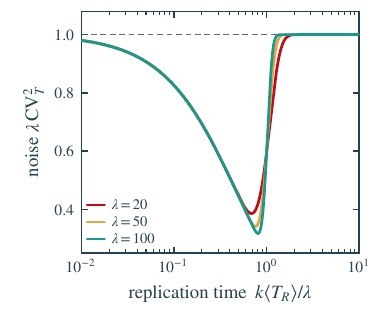}
        \label{fig:ninf}}\hfil
        \subfloat[]{\includegraphics[width=0.32\textwidth]{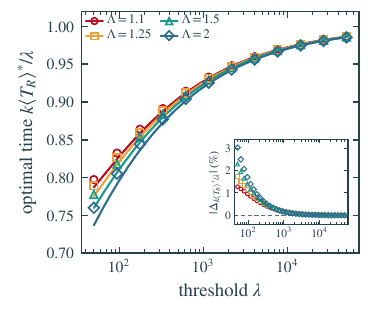}
        \label{fig:optimal_tr}}
        \caption{
        \textbf{The noise in the first passage time $T$ with threshold $\lambda$.} (i) Panels (a) $\lambda=20$, (b) $\lambda=50$, and (c) $\lambda=100$, with $N=1,\dots,10$ (colour scale) at $\Lambda=2$. The noise is normalised by the counting limit $1/\lambda$ (dashed) and shown versus $k\langle T_R\rangle/\lambda$, which equals $\tau$ [Eq.~\eqref{eq:slope}] at $\Lambda=2$. The counting limit is recovered as $\langle T_R\rangle\to 0^+$ and $\langle T_R\rangle\to \infty$. Hence, gene dosage can only reshape the noise at intermediate replication times: it is always reduced below the limit for sufficiently small $\langle T_R\rangle$ and, when replication timing is imprecise (small $N$) and $\Lambda>\Lambda_c$, also amplified above it. (ii) Critical dosage $\Lambda_c$ versus $N/\lambda$, computed from the exact moments [Supplemental Material]: a peak above $1/\lambda$ occurs if and only if $\Lambda>\Lambda_c$. The curves for $\lambda=20$, $50$, and $100$ nearly coincide, with no peak found numerically for $N\ge\lambda$ (dotted line). Shaded band: $\Lambda_c<2$, implying that a peak occurs without dosage compensation ($\Lambda=2$), as in (i), which requires $N/\lambda\lesssim 0.25$. The scale on the vertical axis is logarithmic in $\Lambda_c-1$. (iii) Noise $\lambda\,\CV_T^2$ versus $k\langle T_R\rangle/\lambda$ for deterministic replication ($N\to\infty$), $\Lambda=2$, and $\lambda=20$, $50$, and $100$: only the minimum persists, and timing noise cannot be amplified. (iv) Optimal replication time $k\langle T_R\rangle^*/\lambda$ versus $\lambda$ for $N\to\infty$ and $\Lambda=1.1$, $1.25$, $1.5$, and $2$: numerical computation (symbols) and asymptotic approximation, Eq.~\eqref{eq:optimal_time} (lines); inset: relative error $\Delta_{k\langle T_R\rangle^\ast/\lambda}$ of approximation against computation (absolute value, in percent).}
     \label{fig:results}
    \end{figure*}

\emph{Noise below the counting limit}---Differentiation of Eqs.~\eqref{eq:mean} and \eqref{eq:second} gives, for $\Lambda >1$ and $\langle T_R\rangle\to0^+$,
\begin{equation}
   \lim_{\tau\to 0^+}\frac{d(\lambda\,\CV_T^2)}{d\tau}=-2,
   \quad\text{with}\
   \tau\equiv\frac{(\Lambda-1)k\langle T_R\rangle}{\lambda};
\label{eq:slope}
\end{equation}
see the Supplemental Material.
Here, $(\Lambda-1)k\langle T_R\rangle$ is the mean shortfall in molecules due to replication occurring at $\langle T_R\rangle$ rather than at $t=0$; correspondingly, $\tau$ encodes this shortfall as a fraction of the threshold $\lambda$. Beginning from $\CV_T^2(0^+)=1/\lambda$, the timing noise thus \emph{decreases} as the mean replication time increases, for any $\Lambda>1$ and independently of $N$, as seen for small $\tau$ in Fig.~\ref{fig:cv2}; the initial slope depends on parameters through $\tau$ only. Since $\CV_T^2$ is continuous, returning to $1/\lambda$ as $\langle T_R\rangle\to\infty$, it must attain an interior minimum below the counting limit. That is our central result; see Supplemental Material Theorem III.6: for any threshold $\lambda\ge1$, any dosage factor $\Lambda>1$, and any number of replication steps $N\ge1$, $\CV_T^2$ attains its minimum at some mean replication time $\langle T_R\rangle^{\ast}$ in $(0,\infty)$, with
    \begin{equation}
	   \CV_T^2\big(\langle T_R\rangle^{\ast}\big)<\frac{1}{\lambda}.
	   \label{eq:main}
    \end{equation}
The coupling of accumulation to replication hence makes the threshold crossing \emph{more} precise than molecular counting alone allows.

\emph{Noise above the counting limit}---The same coupling can also make timing less precise than counting alone (Fig.~\ref{fig:cv2}). At fixed $\langle T_R\rangle$, the excess noise $\CV_T^2-1/\lambda$ vanishes at, and is negative just above, $\Lambda=1$, changing sign at most once as $\Lambda$ increases; see the Supplemental Material. Each replication time $\langle T_R\rangle$ therefore has its own threshold dosage above which the noise exceeds $1/\lambda$. Defining the critical dosage $\Lambda_c(\lambda,N)$ as the lowest of these thresholds over all $\langle T_R\rangle$, we show that a peak above $1/\lambda$ occurs if and only if $\Lambda>\Lambda_c$; for $\Lambda\le\Lambda_c$, the noise never exceeds the counting limit. We prove that $\Lambda_c$ is finite whenever $N<\lambda$; see the Supplemental Material. Numerically, $\Lambda_c$ depends mainly on the ratio $N/\lambda$, decreasing only weakly with $\lambda$ for fixed $N/\lambda$ and increasing steeply as $N\to\lambda$. Fig.~\ref{fig:phase} suggests that, for $\lambda = 20$, $50$, and $100$, a peak at $\Lambda = 2$ appears only when $N/\lambda \lesssim 0.25$. Since the noise always dips below $1/\lambda$ at some $\langle T_R\rangle$, reduction and amplification both occur at different replication times.

\emph{Mechanism}---As replication becomes more regular, the peaks shrink and eventually vanish, as shown in Fig.~\ref{fig:cv2}(a), whereas the trough below $1/\lambda$ persists for every $N$. Reduction, amplification, and this disappearance of peaks are best understood from deterministic replication, with $N\to\infty$ at fixed $\langle T_R\rangle$, where the FPT reduces to a simple transformation of the unperturbed accumulation time $X\sim\mathrm{Erlang}(\lambda,k)$:
\begin{equation}
	Y=
	\begin{cases}
		X & \text{if}\ X\le\langle T_R\rangle,\\[4pt]
		\displaystyle \langle T_R\rangle+\frac{X-\langle T_R\rangle}{\Lambda}
		& \text{if}\ X>\langle T_R\rangle.
	\end{cases}
	\label{eq:Ninf}
\end{equation}
Trajectories that reach the threshold before replication ($X\le\langle T_R\rangle$) are unaffected by the increase in gene dosage, whereas the \emph{excess} time of slower trajectories ($X>\langle T_R\rangle$) is compressed by a factor of $\Lambda$. Since too early or too late replication recovers the counting limit, the optimum $\langle T_R\rangle^{\ast}$ lies in between. For $N\to\infty$, we prove that $\langle T_R\rangle^{\ast}$ is the unique critical point, a global minimum below the counting limit (Fig.~\ref{fig:ninf}): without variability in replication timing, coupling can only reduce noise.

Amplification, in turn, originates in that variability. With strong dosage increase ($\Lambda\gg1$) and early replication, protein is hardly produced before replication but, rather, rapidly afterwards. Hence, $T$ reduces to the sum of the replication delay and a short accumulation period. With the ratio $\tau = \mu_3/\mu_2$ fixed, recall Eq.~\eqref{eq:slope}, $\CV_T^2\to(1/\lambda+\tau^2/N)/(1+\tau)^2$, which interpolates between the counting limit and the noise $1/N$ of the replication clock; see the Supplemental Material. The limit exceeds $1/\lambda$ when $N<\lambda$ and $\tau > \tau_{\rm c}= 2N/(\lambda - N)$: when replication is less precise than accumulation and dominates the timing, a sufficiently large dosage factor therefore produces a peak, which implies that $\Lambda_c$ is finite. Peaks hence reflect variability in replication timing, while troughs correspond to a dosage increase, whose initial effect, Eq.~\eqref{eq:slope}, is independent of $N$.

\emph{Optimal replication time}---Having characterised the noise, we now ask when replication should occur. For deterministic replication ($N\to\infty$), we can approximate the optimal mean replication time for $\lambda\to\infty$ and fixed $\Lambda>1$ as
\begin{equation}
    k\langle T_R\rangle^\ast=\lambda-\sqrt{\lambda\,\ln\!\bigg(\frac{\lambda\Lambda^2}{2\pi}\bigg)}+O\!\big[\sqrt{\lambda}\,(\ln\lambda)^{-3/2}\big];
    \label{eq:optimal_time}
\end{equation}
see the Supplemental Material.
Since the accumulation time without replication has mean $\lambda/k$ and standard deviation $\sqrt{\lambda}/k$, Eq.~\eqref{eq:optimal_time} places the optimal replication time about $\big[\ln(\lambda\Lambda^2/2\pi)\big]^{1/2}$ standard deviations before the mean crossing time without replication. Fig.~\ref{fig:optimal_tr} compares this asymptotic approximation with the numerically obtained optimum, which it misestimates by about 1--3\% for $\lambda\in[50,100]$ (inset).

\emph{Discussion}--- We have investigated how the coupling of DNA replication to protein accumulation can tune noise in the timing of cellular events. Depending on when replication occurs, timing noise is either suppressed or amplified, which cells could in principle exploit either to fine-tune the timing of events or to diversify it across a cell population.

Throughout, we consider the FPT distribution within a single replication cycle, as in eukaryotes; overlapping cycles (multi-fork replication), typical of some fast-growing bacteria, are beyond the scope of this work. Bacteria that replicate once per cell cycle, such as slowly growing \textit{E.~coli}, nonetheless remain within scope. In such bacteria, a gene's replication time is set by its position, since replication proceeds from a single origin, \textit{oriC}, at a nearly fixed rate~\cite{CooperHelmstetter1968}; hence, our model can predict an optimal distance from that origin. Chromosomal position is known to affect expression noise through replication timing, both in theory \cite{Swain2002, peterson2015effects} and experimentally in bacteria \cite{Walker2016} and yeast \cite{Chen2016}. However, these studies concern molecule number statistics rather than first-passage times. Measurements of the timing noise of a fluorescent reporter inserted at different chromosomal positions~\cite{Bryant2014} could test this prediction. Its production rate would also need to be measured at each site, because chromosomal position affects expression both through gene dosage, which is higher near \emph{oriC}~\cite{Sousa1997}, and independent of it~\cite{Bryant2014}. In eukaryotes, by contrast, many origins fire stochastically during S phase, with gene timing following the replication-timing program~\cite{Dileep2018}; hence, relocation between early- and late-replicating domains could provide an analogous test. 

Crucially, replication enters only as a production boost at a random time independent of the protein count. Thus, it acts as a second clock alongside protein counting which can sharpen timing beyond what counting alone allows or, when itself noisy, broaden it, potentially producing the peaks we report. As a boost triggered at a fixed protein count would add no independent timing information, timing noise could never fall below $1/\lambda$, the counting limit for a stable, constitutively expressed protein; see the Supplemental Material. Other count-independent boosts, such as cell-cycle cues or external gene-activating signals, may act similarly. A recent study \cite{Verma2026} times replication by the gene's own copy number; by Supplemental Material Proposition~III.7, such a count-triggered boost can only increase timing noise and hence cannot capture the noise reduction below the counting limit reported here.

We identify two major limitations of our present model. First, we neglect protein degradation within the cell cycle, which is a good approximation when protein half-life is much longer than the mean cell-cycle duration, as is the case for many mammalian proteins~\cite{Schwanhusser2011}. The inclusion of degradation would hence allow for our results to be extended to short-lived proteins.

The second limitation is our assumption of constitutive expression. As expression is typically bursty~\cite{Yu2006}, we also simulated bursty production using the Gillespie algorithm \cite{Gillespie1976} included in the Catalyst library~\cite{Loman2023}; see Supplemental Material Section VII and Fig.~S.1. We find that our conclusions remain qualitatively unchanged; however, larger burst sizes require higher molecule number thresholds for peaks to occur. Troughs persist for any burst size, which is consistent with the mechanism reported here: bursting increases counting noise, which drives reduction, but not replication noise, which drives amplification. 

In conclusion, DNA replication acts as a second, count-independent clock. Its coupling to protein accumulation through gene dosage can push timing precision beyond the counting limit, or amplify noise above it when replication is itself imprecise.

\begin{acknowledgments}
		\emph{Acknowledgments}--- J.D.M.-L. acknowledges financial support from a Ph.D. scholarship provided by the Darwin Trust. R.G. acknowledges financial support from the Leverhulme Trust (Grant No. RPG-2024-082), the Stony Brook Foundation, and the Laufer Center for Physical and Quantitative Biology at Stony Brook University.
\end{acknowledgments}
	
 \clearpage
 
  \begin{center}
    {\large\bfseries Supplemental Material for: The impact of DNA replication
    on the timing of cellular events}\\[0.5em]
    Juan David Marmolejo-Lozano, Nikola Popovi\'c, and Ramon Grima
  \end{center}

    \setcounter{section}{0}
  \setcounter{equation}{0}
  \setcounter{figure}{0}
  \setcounter{table}{0}
  \renewcommand{\thefigure}{S.\arabic{figure}}
  \renewcommand{\theHfigure}{S.\arabic{figure}}
  \renewcommand{\theequation}{S\arabic{equation}}
  \renewcommand{\theHequation}{S\arabic{equation}}
  \setcounter{secnumdepth}{3}                    
  \renewcommand{\thesection}{\Roman{section}}    
  \renewcommand{\thesubsection}{\Alph{subsection}} 

  In this supplement, we follow the order of the main text. We first formulate the Chemical Master Equation (CME) and derive the exact first-passage-time (FPT) distribution in the model, as well as the first two moments thereof. We then prove results on the noise below and above the counting limit, analyze the mechanism behind them through the limit of deterministic replication, and finally derive the optimal mean replication time. Throughout, we focus on the mathematical arguments and their proofs.

\section{Model and exact FPT distribution}

Consider the reaction network
\begin{equation}
\begin{aligned}
G_0 &\xrightarrow{\alpha} G_1 \xrightarrow{\alpha} G_2\xrightarrow{\alpha} G_3 \xrightarrow{\alpha} \cdots \xrightarrow{\alpha} G_{N}, \\
G_{j} &\xrightarrow{k} G_j + P\quad\text{for} \ j \in \{0, \cdots, N-1\}, \\
G_N &\xrightarrow{\Lambda k} G_N + P,
\end{aligned}
\label{eq:react_network}
\end{equation}
where a stable product $P$ is synthesised from a gene $G$ with rate $k$. The gene undergoes $N$ steps before replicating, with integer $N\geq 1$; replication multiplies the synthesis rate $k$ by the gene dosage factor $\Lambda\geq 1$. 

\subsection{The CME and its recursive solution in Laplace space}

In our setting, the CME is given by
\begin{align}
    \partial_t P_{j}(i,t) &= k P_j(i-1,t) - k P_{j}(i,t) - \alpha P_j(i,t) + \alpha P_{j-1}(i,t)\quad\text{for} \ j \in \{0, \cdots, N-1\}, \\
    \partial_t P_{N}(i,t) &=  k \Lambda P_N(i-1,t) - k \Lambda P_N(i,t) + \alpha P_{N-1}(i,t),
\end{align}
where $P_{-1}(i,t)\equiv0$ and $P_j(-1,t)\equiv0$; here, $i=0,1,2,\dots$ is a non-negative integer. Using the Laplace transform, we can express these equations as
\begin{align}
    s \tilde{P}_{j}(i,s) - P_{j}(i,t = 0) &= k \tilde{P}_j(i-1,s) - k \tilde{P}_{j}(i,s) - \alpha \tilde{P}_j(i,s) + \alpha \tilde{P}_{j-1}(i,s)\quad\text{for} \ j \in \{0, \cdots, N-1\}, \\
    s \tilde{P}_{N}(i,s) - P_{N}(i,t = 0) &=  k \Lambda \tilde{P}_N(i-1,s) - k \Lambda \tilde{P}_N(i,s) + \alpha \tilde{P}_{N-1}(i,s).
\end{align}
Initially, the gene is in state $G_0$, with no protein present; hence, $P_j(i,t=0)=\delta_{i,0}\delta_{j,0}$, where $\delta_{i,j}$ is the Kronecker delta. In particular, $P_N(i,t=0)=0$ for all $i=0,1,2,\dots$. Hence, we can write
\begin{align}
    s \tilde{P}_{j}(i,s) - \delta_{i,0}\delta_{j,0} &= k \tilde{P}_j(i-1,s) - k \tilde{P}_{j}(i,s) - \alpha \tilde{P}_j(i,s) + \alpha \tilde{P}_{j-1}(i,s)\quad\text{for} \ j \in \{0, \cdots, N-1\}, \\
    s \tilde{P}_{N}(i,s)  &=  k \Lambda \tilde{P}_N(i-1,s) - k \Lambda \tilde{P}_N(i,s) + \alpha \tilde{P}_{N-1}(i,s),
\end{align}
which implies
\begin{align}
    \tilde{P}_{j}(i,s) &= \frac{k}{s+k + \alpha} \tilde{P}_j(i-1,s) + \frac{\alpha}{s+k+\alpha} \tilde{P}_{j-1}(i,s) + \frac{\delta_{i,0}\delta_{j,0}}{s+ k + \alpha}\quad\text{for} \ j \in \{0, \cdots, N-1\}, \\
    \tilde{P}_{N}(i,s)  &=  \frac{k \Lambda}{s+ k\Lambda } \tilde{P}_N(i-1,s) + \frac{\alpha}{s+k \Lambda} \tilde{P}_{N-1}(i,s).
\end{align}
Define
\begin{align}
    a = \frac{k}{s+ k + \alpha},\quad b = \frac{\alpha}{s+k+\alpha},\quad c = \frac{k \Lambda}{s + k \Lambda},\quad\text{and}\quad f = \frac{\alpha}{s + k \Lambda};
    \label{eq:params_Laplace}
\end{align}
then, we can write
\begin{align}
    \tilde{P}_{j}(i,s) &= a \tilde{P}_j(i-1,s) + b\tilde{P}_{j-1}(i,s) + \frac{\delta_{i,0}\delta_{j,0}}{s+ k + \alpha}\quad\text{for} \ j \in \{0,1, \cdots, N-1\}, \label{eq:noburst_rec_1} \\
    \tilde{P}_{N}(i,s)  &=  c \tilde{P}_N(i-1,s) + f \tilde{P}_{N-1}(i,s). \label{eq:noburst_rec_2}
\end{align}
The corresponding initial conditions are given by 
\begin{align}
    \tilde{P}_{0}(0,s) &= \frac{1}{s+k+\alpha}, \label{eq:init_cond_noburst_1}\\
    \tilde{P}_{j}(0,s) &= \tilde{P}_{0}(0,s) b^j = \frac{b^j}{s+k+\alpha}\quad\text{for} \ j \in \{0, \cdots, N-1\}, \label{eq:init_cond_noburst_2}\\
    \tilde{P}_{0}(i,s) &= \tilde{P}_{0}(0,s) a^i = \frac{a^i}{s+k+\alpha}\quad\text{for} \ i \geq 0,\ \text{and}
    \label{eq:init_cond_noburst_3} \\
    \tilde{P}_N(0,s) &= f \tilde{P}_{N-1}(0,s) = \frac{\alpha \ b^{N-1}}{(s+ k \Lambda)(s+k+\alpha)} = \frac{\alpha^N}{(s+ k \Lambda)(s+k+\alpha)^N}. \label{eq:init_cond_noburst_4}
\end{align}
Eqs.~\eqref{eq:init_cond_noburst_2} and \eqref{eq:init_cond_noburst_3} follow by induction, while Eqs.~\eqref{eq:init_cond_noburst_1} and \eqref{eq:init_cond_noburst_4} are obtained by direct substitution.

For the recursion in \eqref{eq:noburst_rec_1}, we propose the solution
\begin{equation}
    \tilde{P}_{j}(i,s) = \frac{a^i b^j}{s+ k + \alpha}\binom{i+j}{j} \quad\text{for} \ i \geq 0 \ \text{and} \ j \in \{0, \cdots N-1\},
    \label{eq:noburst_sol_1}
\end{equation}
which satisfies the initial conditions in Eqs.~\eqref{eq:init_cond_noburst_1} through \eqref{eq:init_cond_noburst_3}. For $(i,j)\neq(0,0)$, substitution into the right-hand side of Eq.~\eqref{eq:noburst_rec_1} gives
\begin{align*}
    a\frac{a^{i-1}b^j}{s+k+\alpha}\binom{i+j-1}{j}
    +b\frac{a^{i}b^{j-1}}{s+k+\alpha}\binom{i+j-1}{j-1}
    &=\frac{a^i b^j}{s+k+\alpha}\left[\binom{i+j-1}{j}+\binom{i+j-1}{j-1}\right]\\
    &=\frac{a^i b^j}{s+k+\alpha}\binom{i+j}{j}=\tilde{P}_j\left(i,s\right),
\end{align*}
where we used Pascal's identity. Hence, Eq.~\eqref{eq:noburst_sol_1} solves the recursion in \eqref{eq:noburst_rec_1}. By \eqref{eq:params_Laplace}, we can equivalently write Eq.~\eqref{eq:noburst_sol_1} as
\begin{equation}
    \tilde{P}_j\left(i,s\right) =  k^i \alpha^j \left(k+s+\alpha\right)^{-\left(1+i+j\right)}\binom{i+j}{j} \quad\text{for} \ i \geq 0 \ \text{and} \ j \in \{0, \cdots, N-1\}.
\end{equation}

Now, to solve the recursion in \eqref{eq:noburst_rec_2}, we note that it is a first-order linear recursion of the form $x(i) = c x(i-1) + g(i)$, with the forcing term $g(i)=f \tilde{P}_{N-1}(i,s)$. One can show that the solution is given by $x(i)=c^i x(0)+\sum_{m=1}^i c^{i-m} g(m)$; then, since $x(0)=\tilde{P}_N(0,s)=f\tilde{P}_{N-1}(0,s)=g(0)$, we can write $x(i)=\sum_{m=0}^{i}c^{i-m}g(m)$. Therefore,
\begin{equation*}
    \tilde{P}_{N}(i,s)  = \frac{f b^{N-1}}{s+k+\alpha}\sum_{m=0}^i c^{i-m} a^m \binom{m+N-1}{m}.
\end{equation*}

Since the protein count can only increase, the probabilities of states with fewer than $\lambda$ molecules are unaffected by absorption at the threshold $\lambda$, for integer $\lambda=1,2,\dots$. The FPT density, which we denote by $h(t)$, is therefore the probability flux from $i=\lambda-1$, the last transient state, into $i=\lambda$; its Laplace transform $\tilde{h}(s)$ reads
\begin{equation*}
    \tilde{h}(s) = k \sum_{j=0}^{N-1} \frac{a^{\lambda-1} b^j}{s+k+\alpha}\binom{\lambda-1+j}{j} + k \Lambda \left(\frac{f b^{N-1}}{s+k+\alpha}\sum_{m=0}^{\lambda-1} c^{(\lambda-1) -m} a^m \binom{m+N-1}{m}\right)
\end{equation*}
or, equivalently,
\begin{align}
\tilde{h}(s) &= k \sum_{j=0}^{N-1} \frac{\big(\frac{k}{s+k+\alpha}\big)^{\lambda-1} \big(\frac{\alpha}{s+k+\alpha}\big)^j}{s+k+\alpha}\binom{\lambda-1+j}{j} \label{eq:FPT_laplace} \\ \nonumber
&\quad+ k \Lambda \left[\frac{\frac{\alpha}{s+ k \Lambda} \big(\frac{\alpha}{s+k+\alpha}\big)^{N-1}}{s+k+\alpha}\sum_{m=0}^{\lambda-1} \left(\frac{k \Lambda}{s+k\Lambda}\right)^{\left(\lambda-1\right) -m} \left(\frac{k}{s+k+\alpha}\right)^m \binom{m+N-1}{m}\right]. 
\end{align}
Importantly, the average time it takes to replicate a gene is given by $\langle T_R \rangle = N/\alpha$, due to the replication time being Erlang distributed. Hence, we can rewrite Eq.~\eqref{eq:FPT_laplace} as
\begin{align}\label{eq:FPT_laplace_2} 
\tilde{h}(s) &= k \sum_{j=0}^{N-1} \frac{\Big(\frac{k}{s+k+N/\langle T_R \rangle}\Big)^{\lambda-1} \Big(\frac{N/\langle T_R \rangle}{s+k+N/\langle T_R \rangle}\Big)^j}{s+k+\frac{N}{\langle T_R\rangle }}\binom{\lambda-1+j}{j} \\ \nonumber
&\quad+ k \Lambda \left[\frac{\frac{N/\langle T_R \rangle}{s+ k \Lambda} \Big(\frac{N/\langle T_R \rangle}{s+k+N/\langle T_R \rangle}\Big)^{N-1}}{s+k+N/\langle T_R \rangle}\sum_{m=0}^{\lambda-1} \left(\frac{k \Lambda}{s+k\Lambda}\right)^{(\lambda-1) -m} \left(\frac{k}{s+k+N/\langle T_R \rangle}\right)^m \binom{m+N-1}{m}\right].
\end{align}

\subsection{Expression of the FPT in terms of special functions}

With $\alpha\equiv N/\langle T_R\rangle$, $A\equiv s+k+\alpha$, and $B\equiv s+k\Lambda$, we can write Eq.~\eqref{eq:FPT_laplace_2} as
\begin{align}\label{eq:FPT_laplace_3}
\tilde{h}(s) &= \left(\frac{k}{A}\right)^\lambda \sum_{j=0}^{N-1} \binom{\lambda + j -1}{j} \left(\frac{\alpha}{A}\right)^j+\left(\frac{\alpha}{A}\right)^N \left(\frac{k \Lambda}{B}\right)^{\lambda} \sum_{m=0}^{\lambda -1} \binom{N+m -1}{m} \left(\frac{B}{A \Lambda}\right)^m \\ \nonumber 
&\equiv S_1+S_2. 
\end{align}

On the other hand, the probability mass function of the negative binomial distribution is given by $P(X = j) = \binom{\lambda+j -1}{j}p^\lambda (1-p)^j$; hence, its cumulative distribution can be written as
\begin{equation}
    \sum_{j=0}^n \binom{\lambda + j -1}{j}p^\lambda (1-p)^j = I_p(\lambda, n+1),
    \label{eq:incomp_Beta}
\end{equation}
where $I_p$ is the incomplete regularised Beta function \cite{NIST:DLMF}. If we define $p$ so that $1-p = \alpha/A$, we have $p^\lambda = [(s+k)/A]^\lambda$, and we can write
\begin{equation}
    \sum_{j=0}^{N-1} \binom{\lambda + j -1}{j}\left(\frac{\alpha}{A}\right)^j = \frac{I_p\left(\lambda, N\right)}{p^\lambda} = \left(\frac{A}{s+k}\right)^{\lambda} I_p(\lambda,N).
\end{equation}
Substituting into $S_1$ in \eqref{eq:FPT_laplace_3}, we obtain
\begin{equation}
    S_1 = \left(\frac{k}{A}\right)^{\lambda}\left(\frac{A}{s+k}\right)^{\lambda} I_p(\lambda, N) = \left(\frac{k}{s+k}\right)^{\lambda} I_p(\lambda,N).
\end{equation}
We now consider the second term $S_2$ in \eqref{eq:FPT_laplace_3}.

Using again the identity in \eqref{eq:incomp_Beta}, with the roles of $\lambda$ and $N$ exchanged and with $p'$ defined by $1-p'=B/(A\Lambda)$, we have $A\Lambda-B=(s+k+\alpha)\Lambda-(s+k\Lambda)=s(\Lambda-1)+\alpha\Lambda$ and
\begin{align}
    p' \equiv \frac{A \Lambda - B}{A \Lambda} = \frac{s(\Lambda -1) + \alpha \Lambda}{(s+k+\alpha) \Lambda},
\end{align}
which implies
\begin{equation}
    \sum_{m=0}^{\lambda-1} \binom{N+m-1}{m}\left(\frac{B}{A \Lambda}\right)^m = \frac{I_{p'}\left(N,\lambda\right)}{(p')^N}.
\end{equation}
Substituting into $S_2$, we find
\begin{equation}
    S_2 = \left(\frac{k \Lambda}{B}\right)^\lambda \left(\frac\alpha{A p'}\right)^N I_{p'}\left(N,\lambda\right)
\end{equation}
Since $(Ap')^N=[(A\Lambda-B)/\Lambda]^N$ implies $[\alpha/(Ap')]^N=[\alpha\Lambda/(A\Lambda-B)]^N$, we obtain
\begin{equation}
    S_2 = \left(\frac{k \Lambda}{s+ k \Lambda}\right)^\lambda \left[\frac{\alpha \Lambda}{s\left(\Lambda -1 \right) + \alpha \Lambda}\right]^N I_{p'}\left(N,\lambda\right)
\end{equation}
Hence, recalling
\begin{equation*}
    p = \frac{s+k}{s+k+N/\langle T_R \rangle}\quad \text{and}\quad p' = \frac{s\left(\Lambda -1 \right) + (N \Lambda)/\langle T_R \rangle}{\left(s+k+N/\langle T_R \rangle\right)\Lambda},
\end{equation*}
we conclude that the FPT distribution in the Laplace space is given by
\begin{equation}
    \boxed{\tilde{h}(s) = \left(\frac{k}{s+k}\right)^\lambda I_p(\lambda, N) + \left(\frac{k \Lambda}{s+ k \Lambda}\right)^\lambda \left[\frac{(N \Lambda)/\langle T_R \rangle}{s(\Lambda -1) + (N\Lambda)/\langle T_R \rangle}\right]^N I_{p'}(N,\lambda).}
    \label{eq:FPT_cute}
\end{equation}

\subsection{Verification of the normalisation property}

At $s=0$, $p_0 = k \langle T_R \rangle/(k \langle T_R \rangle + N)$ and $p_0' = N/(k \langle T_R \rangle + N) = 1 - p_0$.  Therefore, for $s = 0$, the Laplace transform of the FPT distribution is given by 
\begin{equation}
    \tilde{h}(0) =  I_{p_0}(\lambda, N) + I_{p'_0}(N,\lambda),
\end{equation}
with $p_0'=1-p_0$. By the symmetry property of the incomplete regularised Beta function, $I_{1-x}(b,a) = 1 - I_x(a,b)$ \cite{NIST:DLMF}, for $0 \leq x \leq 1$. Thus, $\tilde{h}(0) = 1$, which means that $\tilde{h}\left(s\right)$ is normalised, as required.

\section{Moments of the FPT distribution}

Consider the Laplace transform of the first-passage-time (FPT) distribution given in Eq.~\eqref{eq:FPT_cute}. In this section, we derive closed-form expressions for its mean and second moment.

\subsection{Mean of the FPT distribution}

The mean follows from Eq.~\eqref{eq:FPT_cute}, as $\langle T\rangle=- d\tilde h/ds|_{s=0}$. Applying the product rule, we find

\begin{align*}
    \frac{d\tilde{h}(s)}{ds}\bigg|_{s=0} &= -\frac{\lambda}{k} I_{p_0}(\lambda, N) + \frac{d I_p(\lambda, N)}{ds}\bigg|_{s=0}  - \frac{\lambda}{k \Lambda} I_{p'_0}(N,\lambda) 
    -\frac{(\Lambda -1)\langle T_R \rangle}{\Lambda} I_{p'_0}(N,\lambda) + \frac{d I_{p'}(N,\lambda)}{ds}\bigg|_{s=0}.
\end{align*}
Next, define $\kappa\equiv k+N/\langle T_R\rangle$ and

\begin{equation}
    \phi = \frac{p_0^{\lambda -1 }\left(1-p_0\right)^{N-1}}{\mathcal{B}\left(\lambda,N\right)},
    \label{eq:phi}
\end{equation}
where $\mathcal{B}(\lambda,N)$ is the Beta function. Since $\partial_x I_x(a,b)=x^{a-1}(1-x)^{b-1}/\mathcal{B}(a,b)$ \cite{NIST:DLMF}, $p_0'=1-p_0$, and $\mathcal{B}(N,\lambda)=\mathcal{B}(\lambda,N)$, both regularised incomplete Beta functions $I_p$ and $I_{p'}$ have derivative $\phi$ with respect to their argument at $s=0$, which implies
\begin{equation*}
    \frac{dI_p(\lambda,N)}{ds}\bigg|_{s=0}=\phi\frac{dp}{ds}\bigg|_{s=0}
    =\phi\frac{N/\langle T_R\rangle}{\kappa^2},
    \quad
    \text{and}\quad\frac{dI_{p'}(N,\lambda)}{ds}\bigg|_{s=0}=\phi\frac{dp'}{ds}\bigg|_{s=0}
    =\phi\frac{(\Lambda-1)k-N/\langle T_R\rangle}{\Lambda\kappa^2}.
\end{equation*}
With
\begin{align}
    \mu_1 = \frac{\lambda}{k},\quad \mu_2 = \frac{\lambda}{k \Lambda}, \quad\text{and}\quad \mu_3 = \frac{\left(\Lambda-1\right)\langle T_R \rangle}{\Lambda},
    \label{eq:defn_mu_l}
\end{align}
we therefore obtain
\begin{equation*}
    \langle T\rangle=-\left.\frac{d\tilde h}{ds}\right|_{s=0}
    =\mu_1 I_{p_0}(\lambda,N)+(\mu_2+\mu_3)I_{p_0'}(N,\lambda)
    -\phi\left(\frac{N/\langle T_R\rangle}{\kappa^2}
    +\frac{(\Lambda-1)k-N/\langle T_R\rangle}{\Lambda\kappa^2}\right).
\end{equation*}
The bracketed term simplifies to $(\Lambda-1)/(\Lambda\kappa)$. Defining
\begin{equation}
    C_1\equiv-\frac{\Lambda-1}{\Lambda\kappa}
    =-\frac{(\Lambda-1)\langle T_R\rangle}{\Lambda(k\langle T_R\rangle+N)}
\end{equation}
and using $I_{p_0'}(N,\lambda)=1-I_{p_0}(\lambda,N)$, we find that the mean is given by
\begin{equation}
    \boxed{\langle T \rangle = \mu_1 I_{p_0}(\lambda, N) + (\mu_2 + \mu_3)(1 - I_{p_0}(\lambda, N)) + C_1 \phi.}
    \label{eq:T_mean}
\end{equation}

\subsection{Second moment of the FPT distribution}

Since $\langle T^2\rangle=d^2\tilde h/ds^2|_{s=0}$, we differentiate Eq.~\eqref{eq:FPT_cute} twice. Applying again the product rule, we find

\begin{align*}
   \frac{d^2 \tilde{h}(s)}{ds^2}\bigg|_{s=0}
&= \frac{\lambda(1+\lambda)}{k^2} I_{p_0}(\lambda,N)
- 2\frac{\lambda}{k}\frac{dI_p(\lambda,N)}{ds}\bigg|_{s=0}
+ \frac{d^2 I_p(\lambda,N)}{ds^2}\bigg|_{s=0} + \frac{\lambda(1+\lambda)}{k^2 \Lambda^2}I_{p'_0}(N,\lambda) \\[6pt]
&\quad+ 2\frac{\lambda}{k \Lambda}
\frac{(\Lambda -1)\langle T_R \rangle}{\Lambda}
I_{p'_0}(N,\lambda) - 2\frac{\lambda}{k \Lambda}
\frac{dI_{p'}(N,\lambda)}{ds}\bigg|_{s=0} \\[6pt]
&\quad+ \frac{(1+N)(\Lambda -1)^2\langle T_R \rangle^2}{N\Lambda^2}
I_{p'_0}(N,\lambda) - 2
\frac{(\Lambda -1)\langle T_R \rangle}{\Lambda}
\frac{dI_{p'}(N,\lambda)}{ds}\bigg|_{s=0} + 
\frac{d^2 I_{p'}(N,\lambda)}{ds^2}\bigg|_{s=0}.
\end{align*}
Given that
\begin{equation*}
    \frac{dp}{ds} = \frac{N /\langle T_R \rangle }{(s + \kappa)^2}\quad\text{and}\quad \frac{dp'}{ds} = \frac{k(\Lambda-1) - N/\langle T_R \rangle}{\Lambda(s+ \kappa)^2},
\end{equation*}
where $\kappa=k+N/\langle T_R \rangle$, as before, we find the corresponding second derivatives as
\begin{equation*}
    \frac{d^2p}{ds^2} \bigg|_{s=0} = -\frac{2 N/\langle T_R \rangle}{\kappa^3}\quad\text{and}\quad\frac{d^2 p'}{ds^2}\bigg|_{s=0} = -\frac{2[k(\Lambda-1) - N/\langle T_R \rangle]}{\Lambda \kappa^3}.
\end{equation*}
Moreover, writing $\phi(s)=p^{\lambda-1}(1-p)^{N-1}/\mathcal{B}(\lambda,N)$, which then implies $\phi=\phi(0)$, we have
\begin{equation*}
    \frac{d^2 I_p(\lambda,N)}{ds^2} = \frac{d\phi}{ds}\frac{dp}{ds} + \phi(s) \frac{d^2p}{ds^2}\quad\text{and}\quad
    \frac{d\phi}{ds} = \phi(s)\left(\frac{\lambda-1}{p} - \frac{N-1}{1-p}\right)\frac{dp}{ds};
\end{equation*}
an analogous expression can be obtained for $(d^2 I_{p'}(N,\lambda))/ds^2$, with the roles of $\lambda$ and $N$ exchanged. Hence,
\begin{align*}
   \frac{d^2 I_p(\lambda,N)}{ds^2}\bigg|_{s=0} &= \phi\left[\frac{\lambda -1 }{p_0} - \frac{N-1}{1-p_0}\right]\left(\frac{dp}{ds}\bigg|_{s=0}\right)^2 + \phi \frac{d^2p}{ds^2}\bigg|_{s=0}\quad\text{and} \\
    \frac{d^2 I_{p'}(N,\lambda)}{ds^2}\bigg|_{s=0} &= \phi\left[\frac{N -1}{1-p_0} - \frac{\lambda-1}{p_0}\right]\left(\frac{dp'}{ds}\bigg|_{s=0}\right)^2 + \phi \frac{d^2p'}{ds^2}\bigg|_{s=0},
\end{align*}
which allows us to write
\begin{align*}
 \frac{d^2 \tilde{h}(s)}{ds^2}\bigg|_{s=0}
&= \frac{\lambda(1+\lambda)}{k^2} I_{p_0}(\lambda,N)
  - \frac{2\lambda}{k}\phi\frac{N/\langle T_R\rangle}{\kappa^2} 
  + \phi\left[\frac{\lambda-1}{p_0}- \frac{N-1}{1-p_0}\right]
       \left(\frac{N/\langle T_R\rangle}{\kappa^2}\right)^{2}
  - \frac{2\phi N/\langle T_R\rangle}{\kappa^3} \\[6pt]
&\quad+ \frac{\lambda(1+\lambda)}{k^2\Lambda^2}I_{p'_0}(N,\lambda)
+ \frac{2\lambda\left(\Lambda-1\right)\langle T_R\rangle}{k\Lambda^2}
       I_{p'_0}(N,\lambda) - \frac{2\lambda}{k\Lambda}\phi
       \frac{(\Lambda-1)k - N/\langle T_R\rangle}
            {\Lambda\kappa^2} \\[6pt]
&\quad+ \frac{(1+N)(\Lambda-1)^2\langle T_R\rangle^2}
            {N\Lambda^2}I_{p'_0}(N,\lambda) - \frac{2\left(\Lambda-1\right)\langle T_R\rangle}{\Lambda}\phi
       \frac{(\Lambda-1)k - N/\langle T_R\rangle}
            {\Lambda\kappa^2} \\[6pt]
&\quad+ \phi\left[\frac{N-1}{1-p_0} - \frac{\lambda-1}{p_0}\right]
       \left[\frac{(\Lambda-1)k - N/\langle T_R\rangle}{\Lambda\kappa^2}\right]^{2}
  - \frac{2\phi\left[(\Lambda-1)k
       - N/\langle T_R\rangle\right]}{\Lambda\kappa^3}.
\end{align*}
Defining $\langle t^2 \rangle_{(1)} = \mu_1^2 (1 + 1/\lambda)$, $\langle t^2 \rangle_{(2)} = \mu_2^2 (1 + 1/\lambda)$, and
$\langle t^2 \rangle_{(3)} = \mu_3^2 (1 + 1/N)$, with $\mu_l$ ($l=1,2,3$) as defined in \eqref{eq:defn_mu_l}, and noting that $2\lambda(\Lambda-1)\langle T_R\rangle/(k\Lambda^2)=2\mu_2\mu_3$ and
$(1+N)(\Lambda-1)^2\langle T_R\rangle^2/(N\Lambda^2)=\langle t^2\rangle_{(3)}$, we can write
\begin{align*}
 \frac{d^2 \tilde{h}(s)}{ds^2}\bigg|_{s=0}
&= \langle t^2\rangle_{(1)} I_{p_0}(\lambda,N)
+ \left(\langle t^2\rangle_{(2)} + 2 \mu_2 \mu_3 + \langle t^2\rangle_{(3)}\right) I_{p'_0}(N,\lambda)\\
&\quad+ \phi\left\{ -\frac{2\lambda}{k}\frac{N/\langle T_R\rangle}{\kappa^2}
  + \left[\frac{\lambda-1}{p_0} - \frac{N-1}{1-p_0}\right]\left(\frac{N/\langle T_R\rangle}{\kappa^2}\right)^{2}
  - \frac{2N/\langle T_R\rangle}{\kappa^3}\right. \\
&\quad - \frac{2\lambda}{k\Lambda}\frac{(\Lambda-1)k - N/\langle T_R\rangle}{\Lambda\kappa^2}
  - \frac{2(\Lambda-1)\langle T_R\rangle}{\Lambda}\frac{(\Lambda-1)k - N/\langle T_R\rangle}{\Lambda\kappa^2} \\
  &\quad+ \left.\left[\frac{N-1}{1-p_0} - \frac{\lambda-1}{p_0}\right]\left[\frac{(\Lambda-1)k - N/\langle T_R\rangle}{\Lambda\kappa^2}\right]^{2}
  - \frac{2\left[(\Lambda-1)k - N/\langle T_R\rangle\right]}{\Lambda\kappa^3}\right\}.
\end{align*}
A direct simplification shows that the expression in curly brackets multiplying $\phi$ equals
\begin{align}
    C_2= -\frac{(\Lambda -1) \langle T_R \rangle Q}{k\Lambda^2 N(N + k \langle T_R \rangle)^2},
\end{align}
with
\begin{align}\label{eq:defn_Q}
Q &= \Lambda\left[N^2 (k \langle T_R \rangle + \lambda + 1) + k N \langle T_R \rangle (k \langle T_R \rangle + \lambda) + k^2 \langle T_R \rangle^2 \right] \\ \nonumber
&\quad - (N + k \langle T_R \rangle)[k \langle T_R \rangle (N + 1) - N (\lambda +1)].
\end{align}
With $I_{p_0'}(N,\lambda)=1-I_{p_0}(\lambda,N)$, we therefore have
\begin{equation}
 \boxed{\langle T^2 \rangle = \langle t^2\rangle_{(1)} I_{p_0}(\lambda,N) + \left(\langle t^2\rangle_{(2)} + 2 \mu_2 \mu_3 + \langle t^2 \rangle_{(3)}  \right)(1-I_{p_0}(\lambda, N)) + C_2 \phi.}
\end{equation}
Note that $\mu_1$, $\mu_2$, and $\mu_3$ are the means of the distributions $\mathrm{Erlang}(\lambda,k)$, $\mathrm{Erlang}(\lambda,k\Lambda)$, and $\mathrm{Erlang}\bigl(N,\allowbreak N\Lambda/((\Lambda-1)\langle T_R\rangle)\bigr)$, respectively, whose squared coefficients of variation are $\CV_1^2=\CV_2^2=1/\lambda$ and $\CV_3^2=1/N$; accordingly, $\langle t^2\rangle_{(l)}=\mu_l^2(1+\CV_l^2)$ ($l=1,2,3$).

\subsection{Coefficient of variation of the FPT distribution}

Given the moments $\langle T \rangle$ and $\langle T^2 \rangle$, the definition $\CV_X^2 = \langle X^2 \rangle/\langle X \rangle^2 - 1$ yields a closed-form expression for the squared coefficient of variation of the FPT distribution:
\begin{equation}
    \boxed{\CV^2_T = \frac{\langle t^2\rangle_{(1)} I_{p_0}(\lambda,N) + \left(\langle t^2\rangle_{(2)} + 2 \mu_2 \mu_3 + \langle t^2 \rangle_{(3)}  \right) (1-I_{p_0}(\lambda, N)) + C_2 \phi }{[\mu_1 I_{p_0}(\lambda, N) + (\mu_2 + \mu_3)(1- I_{p_0}(\lambda, N)) + C_1 \phi]^2} -1.}
    \label{eq:noise}
\end{equation}
In the limits of $\langle T_R\rangle\to0^+$ and $\langle T_R\rangle\to\infty$, the FPT distribution tends to $\mathrm{Erlang}(\lambda,k\Lambda)$ and
$\mathrm{Erlang}(\lambda,k)$, respectively, which implies that $\CV_T^2\to1/\lambda$ in both limits; see also Lemma~\ref{lem:tr_ends} below. Any departure from the counting limit $1/\lambda$ must therefore occur at intermediate mean replication times.

Having obtained the moments of the distribution and a closed-form expression for the squared coefficient of variation, we now use those to prove several properties of the timing noise as a function of the mean replication time $\langle T_R\rangle$. 

We first show that, for any $\lambda\ge1$, $N\ge1$, and $\Lambda>1$, there exists a mean replication time at which the noise falls below the counting limit $1/\lambda$, which cannot happen if production depends only on the current protein count. For finite $N$, we prove that, at fixed $\langle T_R\rangle$, the excess noise $\CV_T^2-1/\lambda$ changes sign at most once as the dosage factor $\Lambda$ grows, from negative to positive. It follows that, for all $\lambda$ and $N$, the noise exceeds the counting limit at some $\langle T_R\rangle$ if and only if $\Lambda$ exceeds a critical dosage $\Lambda_c$, and that $\Lambda_c$ is finite whenever $N<\lambda$. Finally, we show that, for deterministic replication ($N\to\infty$), the noise has a single critical point, a global minimum, which implies that no noise amplification (``peak") can occur in that limit.

\section{Noise below the counting limit}

With a constant production rate, the time it takes to accumulate $\lambda$ molecules is Erlang of shape $\lambda$; correspondingly, the timing noise equals the counting limit $1/\lambda$. Coupling accumulation to replication, we show that the noise necessarily falls below that limit at some mean replication time $\langle T_R\rangle$, for every $\Lambda>1$. For both $\langle T_R\rangle\to0^+$ and $\langle T_R\rangle\to\infty$, the noise returns to $1/\lambda$ (Lemma~\ref{lem:tr_ends}); as its initial slope is strictly negative (Lemma~\ref{lem:slope} and Corollary~\ref{cor:tau}); continuity then forces an interior minimum below $1/\lambda$ (Theorem~\ref{thm:below}). Finally, we show that this reduction cannot arise if production depends only on the protein count (Proposition~\ref{prop:count}); rather, it relies on replication occurring at a time that is independent of that count.

\begin{lemma}\label{lem:tr_ends} 
There holds $\CV_T^2\to1/\lambda$ both as $\langle T_R\rangle\to0^{+}$ and as $\langle T_R\rangle\to\infty$, for every $\lambda\ge1$, $N\ge1$, and $\Lambda>1$.
\end{lemma}

\begin{proof}
Consider the FPT distribution $\tilde{h}(s)$ in the Laplace space in the form shown in Eq.~\eqref{eq:FPT_laplace_2}. Taking the limit of $\langle T_R \rangle \to \infty$, we find
\begin{align*}
\lim_{\langle T_R \rangle \to \infty}\tilde{h}(s) = \left(\frac{k}{s+k}\right)^{\lambda}
\end{align*} 
as the only term that remains is the one obtained for $j=0$ in the first sum in Eq.~\eqref{eq:FPT_laplace_2}, while all other terms tend to $0$. Similarly, in the limit as $\langle T_R \rangle \to 0^+$, the term corresponding to $m=0$ in the second sum yields 
\begin{align*}
\lim_{\langle T_R \rangle \to 0^+}\tilde{h}(s) = \left(\frac{k \Lambda}{s+ k \Lambda}\right)^{\lambda}.
\end{align*}
These two limits are the Laplace transforms of $\text{Erlang}(\lambda, k)$ and $\text{Erlang}(\lambda, k\Lambda)$, respectively; hence, by the continuity theorem for Laplace transforms \cite{Feller1966}, the FPT distribution tends to the corresponding Erlang distributions in those two limits.

Convergence in distribution alone does not imply convergence of moments; therefore, we still need to bound the moments of $T$ uniformly in $\langle T_R\rangle$. Recall that $T_R\sim\mathrm{Erlang}(N,N/\langle T_R\rangle)$ denotes the replication time at which a gene initiated in state $G_0$ at $t=0$ reaches state $G_N$, see the main text. Assuming that zero proteins are present initially, we can then write the protein count $n(t)$ as a time-inhomogeneous Poisson process with rate $k$ on $[0,T_R)$ and rate $k\Lambda$ on $[T_R,\infty)$:
\[
n(t)=\Pi_1(t)+\Pi_2((t-T_R)^+).
\]
Here, $\Pi_1$ and $\Pi_2$ are independent Poisson processes with rates $k$ and $k(\Lambda-1)>0$, respectively, that are independent of $T_R$, and $(\cdot)^{+} \equiv \max(\cdot,0)$. Conditionally on $T_R$, the FPT is then $T=\inf\{t: n(t)\ge\lambda\}$. 

Since $0\leq (t-T_R)^+\leq t$, $\Pi_2(0)=0$, and $\Pi_2$ is non-decreasing, we have $\Pi_1(t)\leq n(t)\leq\Pi_1(t)+\Pi_2(t)$ for all $t\geq 0$. Therefore, $T_2\leq T\leq T_1$, where
\[
T_1=\inf\{t: \Pi_1(t)\geq\lambda\}\sim\mathrm{Erlang}(\lambda,k)\quad\text{and}\quad
T_2=\inf\{t: \Pi_1(t)+\Pi_2(t)\geq\lambda\}\sim\mathrm{Erlang}(\lambda,k\Lambda)
\]
for all $\langle T_R\rangle$; note that $\langle T_1\rangle=\mu_1$ and $\langle T_2\rangle=\mu_2$, with $\mu_l$ ($l=1,2$) defined as in \eqref{eq:defn_mu_l}. As $T_1$, $T$, and $T_2$ are non-negative, we have $T_2^m\leq T^m\leq T_1^m$ for any $m\geq 1$ and, hence, $\langle T_2^m\rangle\leq\langle T^m\rangle\leq\langle T_1^m\rangle$. In particular, since the $m$-th moment of $\mathrm{Erlang}(\lambda,k)$ equals $\Gamma(\lambda+m)/(\Gamma(\lambda)k^m)$, it follows that
\[
\sup_{\langle T_R\rangle\in (0,\infty)}\langle T^3\rangle\leq\langle T_1^3\rangle=\frac{\lambda(\lambda+1)(\lambda+2)}{k^3}<\infty.
\]

As the third moment $\langle T^3\rangle$ of $T$ is hence uniformly bounded, convergence in distribution to $\mathrm{Erlang}(\lambda,k)$ as $\langle T_R\rangle\to\infty$ and to $\mathrm{Erlang}(\lambda,k\Lambda)$ as $\langle T_R\rangle\to 0^+$ implies convergence of the first and second moments $\langle T\rangle$ and $\langle T^2\rangle$ in those limits \cite{Billingsley1995}. Specifically, $\langle T\rangle\to\lambda/k>0$ and $\langle T^2\rangle\to\lambda(\lambda+1)/k^2$ as $\langle T_R\rangle\to\infty$, while
$\langle T\rangle\to\lambda/(k\Lambda)>0$ and $\langle T^2\rangle\to\lambda(\lambda+1)/(k\Lambda)^2$ as $\langle T_R\rangle\to 0^+$, respectively, which implies
\begin{equation*}
\CV_T^2=\frac{\langle T^2\rangle}{\langle T\rangle^2}-1\to
\frac{\lambda(\lambda+1)}{\lambda^2}-1=\frac{1}{\lambda}
\end{equation*}
in both limits.
\end{proof}

\begin{corollary}\label{cor:mean_bounds}
There holds $\mu_2\le\langle T\rangle\le\mu_1$ for all $\langle T_R\rangle$; in particular, $\langle T\rangle>0$.
\end{corollary}

\begin{proof}
As $T_2\leq T\leq T_1$, by the proof of Lemma~\ref{lem:tr_ends}, we may take expectations to find

\[
0<\mu_2=\langle T_2\rangle\leq\langle T\rangle\leq\langle T_1\rangle=\mu_1.
\]

\end{proof}

\begin{lemma}\label{lem:slope}
For all $\lambda\ge1$, $N\ge1$, $\Lambda>1$, and $k>0$, there holds
\begin{equation}
    \lim_{\langle T_R\rangle\to 0^{+}}\frac{d\CV_T^2}{d\langle T_R\rangle}
    =-\frac{2k(\Lambda-1)}{\lambda^2}<0.
    \label{eq:noise_derivative}
\end{equation}
\end{lemma}
\begin{proof}
We consider the case of $\lambda>1$ first; the boundary case where $\lambda=1$, in which $\phi$ does not vanish as $\langle T_R\rangle\to0^{+}$, is discussed at the end of the proof.

The derivative of the squared coefficient of variation $\CV_T^2$ is given by
\begin{equation}
   \frac{d\CV_T^2}{d\langle T_R\rangle}
   =\frac{1}{\langle T\rangle^2}\frac{d\langle T^2\rangle}{d\langle T_R\rangle}
   -\frac{2\langle T^2\rangle}{\langle T\rangle^3}\frac{d\langle T\rangle}{d\langle T_R\rangle}.
   \label{eq:dCV2}
\end{equation}
For the derivative of the second moment $\langle T^2\rangle$, we first note that $\langle t^2\rangle_{(1)}$ and $\langle t^2\rangle_{(2)}$ are independent of $\langle T_R\rangle$, and we recall that

\begin{equation}
    \frac{dI_{p_0}(\lambda,N)}{d\langle T_R\rangle}=\phi\frac{dp_0}{d\langle T_R\rangle}=\phi\frac{(1-p_0)p_0}{\langle T_R\rangle}.
\end{equation}
Moreover,
\begin{equation}
    \frac{dQ}{d\langle T_R\rangle}
    =k\left\{N[\lambda\Lambda+2k(\Lambda-1)\langle T_R\rangle+\lambda]
    +2k(\Lambda-1)\langle T_R\rangle+(\Lambda-1)N^2\right\},
\end{equation}
\begin{equation}
    \frac{dC_2}{d\langle T_R\rangle}
    =-\frac{(\Lambda-1)\langle T_R\rangle}{k\Lambda^2N(k\langle T_R\rangle+N)^2}\frac{dQ}{d\langle T_R\rangle}
    -\frac{(\Lambda-1)Q}{k\Lambda^2N(k\langle T_R\rangle+N)^2}
    +\frac{2(\Lambda-1)\langle T_R\rangle Q}{\Lambda^2N(k\langle T_R\rangle+N)^3},
    \label{eq:dC2}
\end{equation}
and
\begin{equation}
    \frac{d\phi}{d\langle T_R\rangle}
    =-\frac{(1-p_0)^{N-1}p_0^{\lambda}[N(1-\lambda)+k\langle T_R\rangle(N-1)]}{k\langle T_R\rangle^2\mathcal{B}(\lambda,N)}
    =-\frac{\phi}{\langle T_R\rangle}[(1-p_0)(1-\lambda)+p_0(N-1)].
\end{equation}

With that in mind, we can write

\begin{align*}
    \frac{d\langle T^2\rangle}{d\langle T_R\rangle}
    &=
    \langle t^2\rangle_{(1)}\frac{dI_{p_0}}{d\langle T_R\rangle}
    +\left(
    2\frac{d\left(\mu_2\mu_3\right)}{d\langle T_R\rangle}
    +\frac{d\langle t^2\rangle_{(3)}}{d\langle T_R\rangle}\right)\left(1-I_{p_0}\right)\\
    &\quad-\left(\langle t^2\rangle_{(2)}+2\mu_2\mu_3+\langle t^2\rangle_{(3)}\right)\frac{dI_{p_0}}{d\langle T_R\rangle}
    +\frac{d\left(C_2\phi\right)}{d\langle T_R\rangle}.
\end{align*}

Moreover, 

$d(2\mu_2\mu_3)/d\langle T_R\rangle=2\mu_2\mu_3/\langle T_R\rangle$ and $d\langle t^2\rangle_{(3)}/d\langle T_R\rangle=2\langle t^2\rangle_{(3)}/\langle T_R\rangle$, which implies
\begin{align*}
    \frac{d\langle T^2\rangle}{d\langle T_R\rangle}
    &=\phi\langle t^2\rangle_{(1)}\frac{dp_0}{d\langle T_R\rangle}
    +\frac{2(1-I_{p_0})}{\langle T_R\rangle}\left(\mu_2\mu_3+\langle t^2\rangle_{(3)}\right)\\
    &\quad-\left(\langle t^2\rangle_{(2)}+2\mu_2\mu_3+\langle t^2\rangle_{(3)}\right)\phi\frac{dp_0}{d\langle T_R\rangle}
    +\frac{dC_2}{d\langle T_R\rangle}\phi+C_2\frac{d\phi}{d\langle T_R\rangle}.
\end{align*}
As $\langle T_R\rangle\to0^+$, $C_2=O(\langle T_R\rangle)$; moreover, for $\lambda>1$, $\phi=O(\langle T_R\rangle^{\lambda-1})$ and
$d\phi/d\langle T_R\rangle=O(\langle T_R\rangle^{\lambda-2})$, so that $C_2d\phi/d\langle T_R\rangle=O(\langle T_R\rangle^{\lambda-1})\to0$; finally, using $Q(\langle T_R\rangle=0)=N^2(\lambda+1)(\Lambda+1)$ in Eq.~\eqref{eq:dC2}, we find that $dC_2/d\langle T_R\rangle\to-(1+\lambda)(\Lambda^2-1)/(kN\Lambda^2)$ remains finite, which implies that $\phi dC_2/d\langle T_R\rangle\to0$ as $\langle T_R\rangle\to 0^+$, as well. Hence, in sum, $d(\phi C_2)/d\langle T_R\rangle\to0$. The remaining terms then reduce, in the limit, to $\lim_{\langle T_R\rangle\to 0^+}2\mu_2\mu_3/\langle T_R\rangle=2\mu_2(\Lambda-1)/\Lambda$, since $\langle t^2\rangle_{(3)}=O(\langle T_R\rangle^{2})$ and $I_{p_0} \to 0$, so that
\begin{equation}
    \lim_{\langle T_R\rangle\to0^{+}}\frac{d\langle T^2\rangle}{d\langle T_R\rangle}
    =\frac{2\lambda(\Lambda-1)}{k\Lambda^2}.
    \label{eq:lem2_lim_1}
\end{equation}

Proceeding similarly for the first moment,
\begin{align*}
\frac{d\langle T\rangle}{d\langle T_R\rangle} &=\frac{d\mu_1}{d\langle T_R\rangle}I_{p_0} +\mu_1\frac{dI_{p_0}}{d\langle T_R\rangle} +\left(\frac{d\mu_2}{d\langle T_R\rangle}+\frac{d\mu_3}{d\langle T_R\rangle}\right)(1-I_{p_0})-(\mu_2+\mu_3)\frac{dI_{p_0}}{d\langle T_R\rangle}+\frac{d(C_1\phi)}{d\langle T_R\rangle}\\
&=\mu_1\phi\frac{dp_0}{d\langle T_R\rangle} +\frac{d\mu_3}{d\langle T_R\rangle}\left(1-I_{p_0}\right) -\phi\frac{dp_0}{d\langle T_R\rangle}\left(\mu_2+\mu_3\right) +\frac{dC_1}{d\langle T_R\rangle}\phi+C_1\frac{d\phi}{d\langle T_R\rangle},
\end{align*}

as $\mu_1$ and $\mu_2$ again do not depend on $\langle T_R\rangle$. As before, $\phi \to 0$ when $\langle T_R\rangle \to 0^+$ for $\lambda>1$, while $C_1=O(\langle T_R\rangle)$ and $d\phi/d\langle T_R\rangle=O(\langle T_R\rangle^{\lambda-2})$, which implies that $C_1d\phi/d\langle T_R\rangle \to 0$; since $dC_1/d\langle T_R\rangle\to-(\Lambda-1)/(\Lambda N)$ and $dp_0/d\langle T_R\rangle \to k/N$ remain finite, it follows that $dC_1/d\langle T_R\rangle \phi \to0$ and $\phi dp_0/d\langle T_R\rangle\to0$, as well. Hence, in particular, $d(C_1\phi)/d\langle T_R\rangle \to 0$. In sum, the only remaining term is $d\mu_3/d\langle T_R\rangle$, whence
\begin{equation}
    \lim_{\langle T_R\rangle\to0^{+}}\frac{d\langle T\rangle}{d\langle T_R\rangle} =\frac{\Lambda-1}{\Lambda}.
\label{eq:lem2_lim_2}
\end{equation}
Substituting Eqs.~\eqref{eq:lem2_lim_1} and \eqref{eq:lem2_lim_2} into Eq.~\eqref{eq:dCV2}, in combination with $\langle T\rangle\to\mu_2=\lambda/(k\Lambda)$ and $\langle T^2\rangle\to\langle t^2\rangle_{(2)}=\mu_2^2(1+1/\lambda)$, we conclude that
\begin{equation}
    \lim_{\langle T_R\rangle\to0^{+}}\frac{d\CV_T^2}{d\langle T_R\rangle}
    =\frac{\frac{2\lambda(\Lambda-1)}{k\Lambda^2}}{\left(\frac{\lambda}{k\Lambda}\right)^2}
    -\frac{2\left(\frac{\lambda}{k\Lambda}\right)^2\left(1+\frac{1}{\lambda}\right)\frac{\Lambda-1}{\Lambda}}
    {\left(\frac{\lambda}{k\Lambda}\right)^3}
    =-\frac{2k(\Lambda-1)}{\lambda^2}.
\end{equation}

Finally, we consider the case where $\lambda=1$; then, $\phi \to N\neq 0$ as $\langle T_R\rangle\to 0^+$. Making use of $p_0 \to 0$, $dp_0/d\langle T_R\rangle \to k/N$, $d\phi/d\langle T_R\rangle \to -(N-1)k$, $C_1,C_2 \to 0$, $dC_1/d\langle T_R\rangle\to-(\Lambda-1)/(\Lambda N)$, and $dC_2/d\langle T_R\rangle\to-2(\Lambda^2-1)/(kN\Lambda^2)$, we obtain
\begin{align*}
    \lim_{\langle T_R\rangle\to0^{+}}\frac{d\langle T^2\rangle}{d\langle T_R\rangle} &=k\left(\langle t^2\rangle_{(1)}-\langle t^2\rangle_{(2)}\right) +\frac{2\mu_2(\Lambda-1)}{\Lambda}-\frac{2(\Lambda^2-1)}{k\Lambda^2} =\frac{2(\Lambda-1)}{k\Lambda^2}\quad\text{and} \\
    \lim_{\langle T_R\rangle\to0^{+}}\frac{d\langle T\rangle}{d\langle T_R\rangle} &=k\mu_1+\frac{\Lambda-1}{\Lambda}-k\mu_2-\frac{\Lambda-1}{\Lambda} =k\left(\mu_1-\mu_2\right)=\frac{\Lambda-1}{\Lambda},
\end{align*}
where in the second line the contribution from $d\mu_3/d\langle T_R\rangle(1-I_{p_0})$ is exactly cancelled by $dC_1/d\langle T_R\rangle \phi $. As these expressions coincide with Eqs.~\eqref{eq:lem2_lim_1} and \eqref{eq:lem2_lim_2}, respectively, evaluated at $\lambda=1$, Eq.~\eqref{eq:noise_derivative} holds for $\lambda=1$, as well.

Finally, since $\CV_T^2$ extends continuously to $\langle T_R\rangle=0$ (Lemma~\ref{lem:tr_ends}), the limit of the
derivative of $\CV_T^2$ as $\langle T_R\rangle\to 0^+$ equals the (one-sided) right derivative at $\langle T_R\rangle=0$.
\end{proof}

\begin{corollary}\label{cor:tau}
Define
\begin{equation}
    \tau\equiv\frac{(\Lambda-1)k\langle T_R\rangle}{\lambda}.
    \label{eq:tau_def}
\end{equation}
Then, for every $\lambda\ge1$, $N\ge1$, $\Lambda>1$, and $k>0$,
\begin{equation}
    \lim_{\tau\to 0^+}\frac{d(\lambda\CV_T^2)}{d\tau}=-2,
    \label{eq:tau_slope}
\end{equation}
which is the form used in the main text.
\end{corollary}

\begin{proof}
For fixed $\lambda$, $N$, $\Lambda$, and $k>0$, $\tau=c\langle T_R\rangle$ with $c\equiv(\Lambda-1)k/\lambda>0$ constant; hence, $\tau\to0^{+}$ if and only if $\langle T_R\rangle\to0^{+}$. Then, by the chain rule,
\begin{equation}
    \frac{d\big(\lambda\CV_T^2\big)}{d\tau}
    =\frac{\lambda}{c}\frac{d\CV_T^2}{d\langle T_R\rangle}.
\end{equation}
Taking the limit as $\tau\to 0^{+}$, we find with Eq.~\eqref{eq:noise_derivative} that
\begin{equation}
    \lim_{\tau\to0^{+}}\frac{d\big(\lambda\CV_T^2\big)}{d\tau}
    =\frac{\lambda^2}{(\Lambda-1)k}\left(-\frac{2k(\Lambda-1)}{\lambda^2}\right)
    =-2.
\end{equation}
\end{proof}

\begin{remark}
By Corollary~\ref{cor:tau}, the initial slope of the normalised noise $\lambda\CV_T^2$ is thus the same for all $\lambda$, $N$ and $\Lambda$, which enter only through $\tau$. To interpret $\tau$, note that the mean number of molecules produced before replication is $k\langle T_R\rangle$.  Had replication occurred at $t=0$, $k\Lambda\langle T_R\rangle$ molecules would have been produced over the interval $[0, \langle T_R \rangle]$. The difference $(\Lambda-1)k\langle T_R\rangle$ is the mean shortfall in molecules caused by replication after time $\langle T_R\rangle$ rather than at the outset; $\tau$ is therefore the shortfall in molecules caused by replication after time $\langle T_R\rangle$ relative to the threshold $\lambda$.
\end{remark}

\begin{theorem}\label{thm:below}
Given any $\lambda\ge1$, $N\ge1$, and $\Lambda>1$, there exists a mean replication time $\langle T_R\rangle^{*}>0$ such that
\begin{equation}
    \CV_T^2(\langle T_R\rangle^{*})<1/\lambda.
\end{equation}
\end{theorem}
\begin{proof}
The moments $\langle T\rangle$ and $\langle T^2\rangle$ are continuous in $\langle T_R\rangle$ on $(0,\infty)$ and $\langle T\rangle>0$ (Corollary~\ref{cor:mean_bounds}), which implies that $\CV_T^2$ is
continuous there. Moreover, by Lemma~\ref{lem:tr_ends}, $\CV_T^2$ extends to a continuous function on the compactified interval $[0,\infty]$, taking the value
$1/\lambda$ at both endpoints. By Lemma~\ref{lem:slope}, the derivative $d\CV_T^2/d\langle T_R\rangle$ is negative on some interval $(0,t_0)$, which implies that $\CV_T^2(t)<1/\lambda$ for $t\in(0,t_0)$. A continuous function on a compact set attains its minimum, which therefore cannot lie at an endpoint; hence, it must be attained at some $\langle T_R\rangle^{*}\in(0,\infty)$ with $\CV_T^2(\langle T_R\rangle^{*})<1/\lambda$.
\end{proof}
In other words, Theorem~\ref{thm:below} states that there exists a mean replication time $\langle T_R\rangle^{*}$ which leads to a decrease in the variability of the time required to reach $\lambda$ protein copies.

Next, we show that the reduction in noise we observe cannot be due to an increase in the production rate of protein alone, but that it necessarily involves DNA replication.
\begin{proposition}[Count-dependent production]\label{prop:count}
Consider a stable (non-degrading) protein with initial count $0$, and let $T$ be the time at which the count first reaches a positive integer threshold $\lambda$. Suppose that the production rate of protein depends only on the current count, taking a value $r_i>0$ while the count equals $i-1$ ($i=1,\dots,\lambda$). Then, $\CV_T^2\ge1/\lambda$, with equality if and only if $r_1=\dots=r_\lambda$.
\end{proposition}
\begin{proof}
Let $\Delta_i$ be the waiting time between the production of the $(i-1)$th and $i$th molecule. While the count is $i-1$, the production rate is a constant $r_i$ that is fixed by the count; hence, $\Delta_i$ is exponentially distributed with rate $r_i$, and $\{\Delta_i\}$ are independent by the Markov property. With $a_i=r_i^{-1}$ and $T=\sum_{i=1}^{\lambda}\Delta_i$, we have
\begin{equation}
    \langle T\rangle=\sum_{i=1}^{\lambda}a_i
    \quad\text{and}\quad
    \mathrm{Var}(T)=\sum_{i=1}^{\lambda}a_i^{2}.
\end{equation}
By the Cauchy-Schwarz inequality, $\big(\sum_{i}a_i\big)^{2}\le\lambda\sum_{i}a_i^{2}$, so
\begin{equation}
    \CV_T^2=\frac{\sum_{i=1}^{\lambda}a_i^{2}}
              {\big(\sum_{i=1}^{\lambda}a_i\big)^{2}}
    \ge \frac{1}{\lambda},
\end{equation}
with equality if and only if all $a_i$ coincide.
\end{proof}
If the dosage increase in our scheme were triggered when the protein count reaches a given value $n$, with $1\leq n<\lambda$, then the production rate would be $r_i=k$ for $i\le n$ and $r_i=k\Lambda$ for $n<i\le\lambda$. Proposition~\ref{prop:count} would then give $\CV_T^2\ge1/\lambda$, with equality holding only for $\Lambda=1$, in contradiction to Theorem~\ref{thm:below}.

\section{Noise above the counting limit}

Having shown that coupling always lowers the noise below the counting limit at some mean replication time, we now conclude that the same coupling can also raise it above that limit; whether it does so is decided, for given $\lambda$ and $N$, by the dosage factor $\Lambda$. 

We say that the noise has a peak when $\CV_T^2$, as a function of the mean replication time $\langle T_R\rangle$ on $(0,\infty)$, attains a local maximum that is strictly larger than $1/\lambda$. Specifically, we prove that there is a critical value $\Lambda_c$, depending only on $\lambda$ and $N$, such that peaks occur precisely for $\Lambda>\Lambda_c$, whereas for $\Lambda\le\Lambda_c$, the noise stays at or below $1/\lambda$ at any $\langle T_R\rangle$. We then show that $\Lambda_c$ is finite whenever $N<\lambda$, that is, whenever replication timing is less reproducible than the accumulation it gates. Reduction and amplification therefore coexist in the same model at different mean replication times. Throughout, we take $\lambda\ge1$, $N\ge1$, $\Lambda\ge1$, $k>0$, and $\langle T_R\rangle>0$, and we write $I\equiv I_{p_0}(\lambda,N)$ for brevity.

Equation~\eqref{eq:FPT_cute} is the sum of two contributions: a pure Erlang term and a mixed term that incorporates the accelerated phase and the replication delay, with mean time scales $\mu_2$ and $\mu_3$, respectively. To isolate the second term, we require $I\to0$ and, hence, $p_0\to0$, which requires $\langle T_R\rangle\to 0^+$. In that limit, however, $\mu_3\to0$, as well, and the mixed term collapses to a pure $\mathrm{Erlang}(\lambda,k\Lambda)$. We therefore take $\langle T_R\rangle\to 0^+$ and $\Lambda\to\infty$ simultaneously, keeping the ratio of the two mean time scales fixed. That ratio is precisely the rescaled delay of Eq.~\eqref{eq:tau_def},
\begin{equation}
    \tau=\frac{\mu_3}{\mu_2}=\frac{\left(\Lambda-1\right)k\langle T_R\rangle}{\lambda}.
\end{equation}
For fixed $\lambda$, $\Lambda>1$, and $k$, the map $\langle T_R\rangle\mapsto\tau$ is a strictly increasing bijection of $\left(0,\infty\right)$ onto itself, with inverse
\begin{equation}
    \langle T_R\rangle_\tau \equiv \frac{\lambda\tau}{k(\Lambda-1)}.
    \label{eq:TRtau}
\end{equation}
Fixing $\tau$ and taking $\Lambda\to\infty$ thus implies $\langle T_R\rangle=\langle T_R\rangle_\tau\to0$ with $p_0\to0$, while $\mu_3/\mu_2$ remains constant by definition.

\begin{lemma}[Cancellation]\label{lem:cancel}
For every $p_0\in(0,1)$, there holds
\begin{equation}
  \lambda I - p_0\phi
  = \frac{N-1}{\mathcal{B}(\lambda,N)}\int_0^{p_0} t^{\lambda}(1-t)^{N-2}dt,
  \label{eq:cancel}
\end{equation}
as well as
\begin{equation}
  0\le\lambda I-p_0\phi\le\frac{(N-1)p_0^{\lambda+1}}{(\lambda+1)\mathcal{B}(\lambda,N)},
  \quad
  I\le\frac{p_0^{\lambda}}{\lambda\mathcal{B}(\lambda,N)},
  \quad\text{and}\quad
  \phi\le\frac{p_0^{\lambda-1}}{\mathcal{B}(\lambda,N)} .
  \label{eq:cancel_bounds}
\end{equation}
\end{lemma}

\begin{proof}
Integration of $I=\mathcal{B}(\lambda,N)^{-1}\int_0^{p_0}t^{\lambda-1}(1-t)^{N-1}dt$ by parts with $u=(1-t)^{N-1}$ and $dv=\lambda t^{\lambda-1}dt$ gives \eqref{eq:cancel}, since the boundary term is exactly $p_0\phi$. For $N=1$, the right-hand side of \eqref{eq:cancel} vanishes; for $N\ge2$, $(1-t)^{N-2}\le1$ gives the first bound. The second follows from $(1-t)^{N-1}\le1$, and the third from the definition of $\phi$; recall Eq.~\eqref{eq:phi}.
\end{proof}

\begin{proposition}\label{prop:limit}
Let $\lambda>1$ and $N\ge1$. Then, for every $\tau>0$,
\begin{equation}
    \lim_{\Lambda\to\infty}
    \CV_T^2\big(\langle T_R\rangle_\tau;\Lambda\big)
    = \mathcal{V}(\tau) \equiv
    \frac{\dfrac{1}{\lambda}+\dfrac{\tau^{2}}{N}}{\left(1+\tau\right)^{2}} .
    \label{eq:Vtau}
\end{equation}
The limit in Eq.~\eqref{eq:Vtau} tends to $1/\lambda$ as $\tau\to0^{+}$ and to $1/N$ as $\tau\to\infty$; moreover, it exceeds $1/\lambda$ if and only if
\begin{equation}
    N<\lambda
    \quad\text{and}\quad
    \tau>\tau_{c}\equiv\frac{2N}{\lambda-N} .
    \label{eq:taustar}
\end{equation}
\end{proposition}
\begin{proof}
Write $x=k\langle T_R\rangle_\tau=\lambda\tau/(\Lambda-1)$, so that $x\to0^+$ as $\Lambda\to\infty$, with $\Lambda=1+\lambda\tau/x$ and $p_0=x/(x+N)=O(x)$ as $x\to 0^+$. Since $\mu_1=\Lambda\mu_2$, $\mu_3=\tau\mu_2$, and $\langle t^2\rangle_{(l)}=\mu_l^2(1+\CV_l^2)$ for $l=1,2,3$, the moments $\langle T\rangle$ and $\langle T^2\rangle$ factor as
\begin{align*}
    \langle T\rangle &= \mu_2\left[\Lambda I+\left(1+\tau\right)\left(1-I\right)
    +\frac{C_1\phi}{\mu_2}\right]\quad\text{and} \\
    \langle T^2\rangle &= \mu_2^{2}\left\{\frac{(\lambda+1)\Lambda^{2}}{\lambda}I
    +\left[\frac{\lambda+1}{\lambda}+2\tau+\frac{\tau^{2}\left(N+1\right)}{N}\right]\left(1-I\right)+\frac{C_2\phi}{\mu_2^{2}}\right\},
\end{align*}
respectively. Hence, $\mu_2$ cancels in $\CV_T^2=\langle T^2\rangle/\langle T\rangle^2-1$; moreover, since $I\to0$, it suffices to show, using Lemma~\ref{lem:cancel}, that $\Lambda I+C_1\phi/\mu_2\to0$ and $(\lambda+1)\Lambda^{2}I/\lambda+C_2\phi/\mu_2^2\to0$ as $\Lambda\to\infty$. While the two terms in each pair do not vanish individually when $x$ is small, their sum does, by Lemma~\ref{lem:cancel}.

Since $C_1=-(\Lambda-1)p_0/(k\Lambda)$ and $\mu_2^{-1}=k\Lambda/\lambda$, we have
$C_1\phi/\mu_2=-(\Lambda-1)p_0\phi/\lambda$ exactly, so that
\begin{equation*}
    \Lambda I+\frac{C_1\phi}{\mu_2}
    =\frac{\Lambda}{\lambda}(\lambda I-p_0\phi)+\frac{p_0\phi}{\lambda}
    =O(\Lambda x^{\lambda+1})+O(x^{\lambda})=O(x^{\lambda}),
\end{equation*}
by \eqref{eq:cancel_bounds} and $\Lambda=O(x^{-1})$.

Expanding $Q$, as defined in \eqref{eq:defn_Q}, at $x=0$, we find $Q=N^{2}(\lambda+1)(\Lambda+1)[1+O(x)]$ and, using $\langle T_R\rangle/(N+x)^{2}=p_0(1-p_0)/(Nk)$,
$C_2=-(\lambda+1)p_0\big(1-\Lambda^{-2}\big)k^{-2}[1+O(x)]$. Since $\mu_2^{-2}=(k\Lambda/\lambda)^2$,
\begin{equation*}
    \frac{(\lambda+1)\Lambda^{2}}{\lambda}I+\frac{C_2\phi}{\mu_2^2}
    =\frac{(\lambda+1)\Lambda^{2}}{\lambda^{2}}
    \bigl\{\lambda I-p_0\phi+p_0\phi\big[\Lambda^{-2}+O(x)\big]\big\}
    =O(x^{\lambda-1}),
\end{equation*}
again by \eqref{eq:cancel_bounds}, which vanishes as $x\to 0^+$ due to $\lambda>1$.

Therefore,
\begin{equation*}
    \lim_{\Lambda\to\infty}\CV_T^2\big(\langle T_R\rangle_\tau;\Lambda\big)=\frac{(\lambda+1)/\lambda+2\tau+\tau^{2}(N+1)/N}{(1+\tau)^2}-1,
\end{equation*}
which gives \eqref{eq:Vtau}. The limits in $\tau$ are immediate. Finally, $\mathcal{V}(\tau)>1/\lambda$ is equivalent to $\tau^{2}/N>\tau(2+\tau)/\lambda$, i.e.\ to $\tau(\lambda-N)>2N$, which requires $\lambda>N$ and $\tau>\tau_c$, where $\tau_c$ is as defined in \eqref{eq:taustar}.
\end{proof}

\begin{lemma}[Peak criterion]\label{lem:criterion}
For $\Lambda>1$, the noise has a peak if and only if $\CV_T^2\left(\langle T_R \rangle;\Lambda\right)>1/\lambda$ for some $\langle T_R\rangle>0$. For $\Lambda=1$, $\CV_T^2\equiv1/\lambda$, which implies that there is no peak.
\end{lemma}
\begin{proof}
Necessity holds by definition. Conversely, let $\eta=\CV_T^2(\langle T_R\rangle,\Lambda)-1/\lambda>0$. By Lemma~\ref{lem:tr_ends}, there exist $0<t_-<\langle T_R\rangle<t_+$ with $\CV_T^2<1/\lambda+\eta/2$ outside $[t_-,t_+]$. As $\CV_T^2$ is continuous on that interval due to $p_0$, $I$, and $\phi$ being continuous in $\langle T_R\rangle$ and $\langle T\rangle>0$ (Corollary~\ref{cor:mean_bounds}), it attains a maximum there. That maximum is at least $1/\lambda+\eta$ and, hence, a global, and therefore also local, maximum on $(0,\infty)$. 

For $\Lambda=1$, we have $\mu_2=\mu_1$ and $\mu_3=C_1=C_2=0$, since $\mu_3$, $C_1$ and $C_2$ are proportional to $\Lambda-1$; hence, $\langle T\rangle=\mu_1$, $\langle T^2\rangle=\langle t^2\rangle_{(1)}$, and $\CV_T^2\equiv 1/\lambda$.
\end{proof}

\begin{lemma}[Dosage dependence]\label{lem:factor}
Fix $\langle T_R\rangle>0$, write $x=k\langle T_R\rangle$, as before, and define $\mathcal{D}(\Lambda)=k^2\big[\langle T^2\rangle-(1+1/\lambda)\langle T\rangle^2\big]$, which implies that $\CV_T^2>1/\lambda$ if and only if $\mathcal{D}>0$. Then, for every $\Lambda\ge1$,
\begin{equation}
  \mathcal{D}(\Lambda)
  = \frac{\Lambda-1}{\Lambda^2}\big[(\Lambda-1)\mathcal{D}_\infty+\mathcal{D}_1\big],
  \quad\text{with}\ 
  \mathcal{D}_1=2x\big[p_0\mathbb{P}(B=N)-\mathbb{P}(B\ge N)\big]<0,
  \label{eq:factor}
\end{equation}
where $\mathcal{D}_\infty=\lim_{\Lambda\to\infty}\mathcal{D}(\Lambda)$ and $B\sim\mathrm{Binom}(\lambda+N-1,1-p_0)$. Consequently, the set of $\Lambda\ge1$ with $\CV_T^2(\langle T_R\rangle;\Lambda)>1/\lambda$ is empty if $\mathcal{D}_\infty\le0$, and equals the open half-line $\bigl(\Lambda^{\dagger}(\langle T_R\rangle),\infty\bigr)$ if $\mathcal{D}_\infty>0$, where $\Lambda^{\dagger}(\langle T_R\rangle)\equiv1-\mathcal{D}_1/\mathcal{D}_\infty$.
\end{lemma} 
\begin{proof}
Since $p_0$, $I$, $\phi$, and $\mu_1$ do not depend on $\Lambda$, while $\mu_2$, $\mu_3$, and $C_1$ are affine and $C_2$ is quadratic in $\Lambda^{-1}$ due to $Q$ being affine in $\Lambda$, $\Lambda^2\mathcal{D}(\Lambda)$ is a polynomial of degree at most two in $\Lambda$ which vanishes at $\Lambda=1$, by Lemma~\ref{lem:criterion}. Hence, we can write $\Lambda^2\mathcal{D}(\Lambda)=(\Lambda-1)\mathcal{P}(\Lambda)$, with $\mathcal{P}$ affine, $\mathcal{P}(1)=\partial\mathcal{D}/\partial\Lambda|_{\Lambda=1}\equiv\mathcal{D}_1$, and slope $\mathcal{D}_\infty$, which is the factorisation in \eqref{eq:factor}.

At $\Lambda=1$, $k\langle T\rangle=\lambda$ and $\mu_3=0$, so $\mathcal{D}_1=k^2\partial\langle T^2\rangle/\partial\Lambda-2(\lambda+1)k\partial\langle T\rangle/\partial\Lambda$. Using $\partial\mu_2/\partial\Lambda|_{\Lambda=1}=-\lambda/k$, $\partial\mu_3/\partial\Lambda|_{\Lambda=1}=\langle T_R\rangle$, $\partial C_1/\partial\Lambda|_{\Lambda=1}=-\langle T_R\rangle/(x+N)$, and $\partial C_2/\partial\Lambda|_{\Lambda=1}=-2\langle T_R\rangle[N(\lambda+1)+\lambda x]/[k(N+x)^2]$ due to $Q|_{\Lambda=1}=2N[N(\lambda+1)+\lambda x]$, one finds
\begin{equation*}
  \mathcal{D}_1=-2x(1-I)+\frac{2x^2}{(N+x)^2}\phi
  =2x\left[\frac{p_0^{\lambda}(1-p_0)^{N}}{N\mathcal{B}(\lambda,N)}-(1-I)\right].
\end{equation*}
By the binomial representation of $I_x(a,b)$ and the relation $I_{1-x}(b,a)=1-I_x(a,b)$, $1-I=\mathbb{P}(B\ge N)$, and since $[N\mathcal{B}(\lambda,N)]^{-1}=\binom{\lambda+N-1}{N}$, the first term in the square bracket equals $p_0\mathbb{P}(B=N)<\mathbb{P}(B\ge N)$. Hence, $\mathcal{D}_1<0$.

Finally, for $\Lambda>1$, $\mathcal{D}>0$ if and only if $(\Lambda-1)\mathcal{D}_\infty>-\mathcal{D}_1$, i.e.\ if and only if $\mathcal{D}_\infty>0$ and $\Lambda>\Lambda^{\dagger}$.
\end{proof}
Note that if $\Lambda = 1$, $\mathcal{D}(1) \equiv 0$; therefore, that point never belongs to the set constructed in Lemma~\ref{lem:factor}.

In sum, we can conclude that, for fixed $\langle T_R\rangle$, the excess noise $\CV_T^2-1/\lambda$ vanishes at $\Lambda=1$, is negative for $\Lambda$ slightly above $1$, and changes sign at most once as $\Lambda$ grows, from negative to positive.

With $\lambda$ and $N$ fixed, whether the noise has a peak depends only on $\Lambda$; hence, we set
\begin{equation}
  \Lambda_c \equiv
  \inf\big\{\Lambda\ge1: \CV_T^2(\cdot;\Lambda)\ \text{has a peak}\big\}
  \in[1,\infty].
  \label{eq:Lambdac}
\end{equation}

\begin{theorem}\label{thm:main}
For every $\lambda,N\ge1$, the noise has a peak if and only if $\Lambda>\Lambda_c$; in particular, for $\Lambda\le\Lambda_c$, $\CV_T^2\le1/\lambda$ for all $\langle T_R\rangle>0$. If, moreover, $N<\lambda$, then $\Lambda_c<\infty$: for every $\tau>\tau_c$, there exists $\Lambda_\tau<\infty$ such that $\CV_T^2\big(\langle T_R\rangle_\tau;\Lambda\big)>1/\lambda$ for every $\Lambda>\Lambda_\tau$, and
\begin{equation}
    \Lambda_c\le\inf_{\tau>\tau_c}\Lambda_\tau<\infty .
\end{equation}
\end{theorem}
\begin{proof}
By Lemma~\ref{lem:criterion}, the set defined in \eqref{eq:Lambdac} is the union over $\langle T_R\rangle>0$ of the sets $\{\Lambda\ge 1 :\CV_T^2(\langle T_R\rangle;\Lambda)>1/\lambda\}$, each of which is either empty or an open half-line, by Lemma~\ref{lem:factor}. A union of open half-lines is the open half-line whose left endpoint is the infimum of the individual endpoints; therefore, the set equals $(\Lambda_c,\infty)$ and is empty if $\Lambda_c=\infty$. For $\Lambda\le\Lambda_c$, there is therefore no peak, and Lemma~\ref{lem:criterion} gives $\CV_T^2\le1/\lambda$ everywhere.

Now let $N<\lambda$, so that $\lambda>1$, and fix $\tau>\tau_c$. By Proposition~\ref{prop:limit}, $\eta\equiv\mathcal{V}(\tau)-1/\lambda>0$, and there is $\Lambda_\tau$ such that, for every $\Lambda>\Lambda_\tau$,
\begin{equation*}
    \CV_T^2\big(\langle T_R\rangle_\tau;\Lambda\big)
    >\mathcal{V}(\tau)-\frac{\eta}{2}=\frac{1}{\lambda}+\frac{\eta}{2}>\frac{1}{\lambda},
\end{equation*}
with $\langle T_R\rangle_\tau>0$ by \eqref{eq:TRtau}. By Lemma~\ref{lem:criterion}, the noise then has a peak for each such $\Lambda$, so $\Lambda_c\le\Lambda_\tau$.
\end{proof}

\begin{remark}\label{rem:Lc}
Note that the limit of $p_0 \to 0$ enters only through Proposition~\ref{prop:limit}, which is used to show that $\Lambda_c<\infty$ when $N<\lambda$; Lemma~\ref{lem:factor} and the first claim of Theorem~\ref{thm:main} hold for every $p_0\in(0,1)$. The limit is enforced to ensure the existence of peaks: fixing $\tau$ and letting $\Lambda \to \infty$ forces $\langle T_R \rangle \to 0^+$ and, hence, $p_0 \to 0$, which means that replication is much faster than protein production. The threshold itself follows from the proof of Theorem ~\ref{thm:main}: $\Lambda_c=\inf\{\Lambda^{\dagger}(\langle T_R\rangle): \langle T_R\rangle>0,\ \mathcal{D}_\infty(\langle T_R\rangle)>0\}$. Each $\Lambda^{\dagger}$ exceeds $1$, but the infimum need not be attained and the argument does not exclude $\Lambda_c=1$. For $N\ge\lambda$, Proposition~\ref{prop:limit} gives $\mathcal{V}(\tau)\le1/\lambda$ for every $\tau$ and $p_0$ small; hence, the fixed-$\tau$ regime produces no peaks, and numerically none occur at any $\Lambda$, i.e.\ $\Lambda_c=\infty$; we do not give a general proof.
\end{remark}

\section{Mechanism}

We have shown that, for every $\lambda \ge 1$ and $N \ge 1$, the noise falls below the counting limit at some $\langle T_R\rangle$ (Theorem~\ref{thm:below}), and that it exceeds that limit for some $\langle T_R\rangle$ if and only if $\Lambda>\Lambda_c$, with $\Lambda_c$ finite whenever $N<\lambda$ (Theorem~\ref{thm:main}). To understand the mechanism behind both troughs and peaks, we first obtain the FPT distribution in the limit of deterministic replication, where $N\to\infty$. We then categorise the contribution of DNA replication and protein production to the reduction or amplification of noise: specifically, we show that noise is reduced by a dosage increase, which is why (unique) troughs are still observed for deterministic replication (Theorem~\ref{thm:unique}); peaks, by contrast, require variability in replication timing and hence vanish in the limit as $N\to\infty$ (Corollary~\ref{cor:nopeaks}).

\subsection{FPT for deterministic replication}

Recall the definitions of $p$ and $p'$, as well as of the distribution $\tilde{h}(s)$, which we restate here for reference, with
\begin{equation*}
p = \frac{s+k}{s+k+N/\langle T_R \rangle}\quad\text{and}\quad p' = \frac{s(\Lambda -1) + (N\Lambda)/\langle T_R \rangle}{(s+k+N/\langle T_R \rangle)\Lambda},
\end{equation*}
as well as
\begin{equation*}
    \tilde{h}(s) = \left(\frac{k}{s+k}\right)^\lambda I_p(\lambda, N) + \left(\frac{k \Lambda}{s+ k \Lambda}\right)^\lambda \left[\frac{(N \Lambda)/\langle T_R \rangle}{s(\Lambda -1) + (N \Lambda)/\langle T_R \rangle}\right]^N I_{p'}(N,\lambda),
\end{equation*}
respectively.

Consider first the regularised incomplete Beta function $I_p(\lambda, N)$ in the limit as $N\to\infty$. By definition, we have that 
\begin{equation}
    I_x(a,b) = \mathbb{P}(B\geq a),\quad\text{with}\ B \sim \text{Binom}(a+b-1,x)
\end{equation}
As $N \to \infty$, $\lambda + N -1 \to \infty$ and $p \to 0$, which is equivalent to the classical limit of a binomial distribution with a very large number of trials and a very low probability of success tending to a Poisson distribution \cite{Feller1968}. Additionally, the mean $(\lambda + N -1)[(s+k)/(s + k + N/\langle T_R \rangle)]$ of $\text{Binom}(\lambda+N-1,p)$ tends to $(s+k)\langle T_R \rangle$ for $N \to \infty$. Hence,
\begin{equation}
    I_p(\lambda,N) \xrightarrow{N \to \infty} \mathbb{P}(Z \geq \lambda),\quad\text{with}\ Z \sim \text{Poisson}((s+k)\langle T_R \rangle).
\end{equation}
We note that $\mathbb{P}(Z \geq \lambda) = 1 - \mathbb{P}(Z \leq \lambda -1)$. Moreover, we have that
\begin{equation}
    \mathbb{P}(Z \leq \lambda -1) = e^{-(s+k)\langle T_R \rangle} \sum_{j=0}^{\lambda-1} \frac{[(s+k)\langle T_R \rangle]^j}{j!} = \frac{\Gamma(\lambda, (s+k)\langle T_R \rangle)}{\Gamma(\lambda)},
\end{equation}
which implies 
\begin{equation}
     I_p(\lambda,N) \xrightarrow{N \to \infty} 1 - \frac{\Gamma(\lambda, (s+k)\langle T_R \rangle)}{\Gamma(\lambda)} = 1- \frac{\Gamma(\lambda, (s+k)\langle T_R \rangle)}{(\lambda-1)!}.
\end{equation}

The limit of $N\to\infty$ in $I_{p'}(N,\lambda)$ can be described in a similar fashion, as $I_{p'}(N,\lambda) = 1- I_{1-p'}(\lambda, N)$: as the mean of $\text{Binom}(\lambda + N -1, 1- p')$ tends to $(s + k \Lambda)\langle T_R \rangle/\Lambda$, $I_{p'}(N, \lambda) \to  \Gamma(\lambda, (s + k \Lambda)\langle T_R \rangle/\Lambda)/(\lambda-1)!$. 

Finally, we consider the prefactor of $I_p'$ in the definition of $\tilde{h}$:
\begin{equation*}
\left[\frac{(N \Lambda)/\langle T_R \rangle}{s(\Lambda -1) + (N \Lambda)/\langle T_R \rangle}\right]^N
=\left[\frac{1}{1+ s(\Lambda -1)\langle T_R \rangle/(N \Lambda)}\right]^N
\xrightarrow{N \to \infty} e^{-\frac{s(\Lambda-1)\langle T_R \rangle}{\Lambda}},
\end{equation*}
where we have used that $(1+y/N)^N \to e^y$ as $N \to \infty$.

In sum, we have shown that, for $N\to \infty$, our distribution tends to
\begin{equation}
    \lim_{N \to \infty}\tilde{h}(s) = \left(\frac{k}{s + k}\right)^{\lambda} \left(1 - \frac{\Gamma(\lambda, \left(s+k\right)\langle T_R \rangle)}{(\lambda-1)!}\right) + \left(\frac{k \Lambda}{s + k \Lambda}\right)^{\lambda}  \frac{\Gamma(\lambda, (s + k \Lambda)\langle T_R \rangle/\Lambda)}{(\lambda-1)!} e^{-\frac{s(\Lambda-1)\langle T_R \rangle}{\Lambda}}.
    \label{eq:Ninf_approx}
\end{equation}
Since the inverse Laplace transform is linear, we invert the two terms in \eqref{eq:Ninf_approx} separately. For the density $f_X$ of Erlang$(\lambda,k)$ and any $a>0$,
\begin{equation}
    \int_{a}^{\infty}e^{-st}f_X(t)dt
    =\left(\frac{k}{s+k}\right)^{\lambda}\frac{\Gamma(\lambda,(s+k)a)}{(\lambda-1)!},
    \label{eq:erlang_tail_LT}
\end{equation}
as follows from the substitution $u=(s+k)t$. With $a=\langle T_R\rangle$, the first term therefore gives
\begin{equation}
    \mathcal{L}^{-1}\left\{\left(\frac{k}{s + k}\right)^{\lambda} \left(1 - \frac{\Gamma(\lambda, (s+k)\langle T_R \rangle)}{(\lambda-1)!}\right)\right\} = \frac{k^{\lambda } e^{-k t} t^{\lambda -1}}{(\lambda -1)!}[1-\Theta (t-\langle T_R\rangle)]\equiv f_1(t), 
\end{equation}
where $\Theta$ denotes the usual Heaviside function.

For the second term, let $k'\equiv k\Lambda$ and $T'\equiv\langle T_R\rangle/\Lambda$, and note that $\langle T_R\rangle-T'=(\Lambda-1)\langle T_R\rangle/\Lambda (=\mu_3)$. We can then write
\begin{equation*}
    \left(\frac{k \Lambda}{s + k \Lambda}\right)^{\lambda}  \frac{\Gamma(\lambda, (s + k \Lambda)\langle T_R \rangle/\Lambda)}{(\lambda-1)!} e^{-\frac{s(\Lambda-1)\langle T_R \rangle}{\Lambda}} =\left(\frac{k'}{s + k'}\right)^{\lambda} \frac{\Gamma(\lambda, (s + k')T')}{(\lambda-1)!} e^{-s(\langle T_R \rangle - T')}.
\end{equation*}
Next, from the inverse Laplace transform of the first term, we deduce
\begin{equation*}
    \mathcal{L}^{-1}\left\{\left(\frac{k'}{s + k'}\right)^{\lambda} \frac{\Gamma(\lambda, (s + k')T')}{(\lambda-1)!}\right\} = \frac{k'^{\lambda } e^{-k' t} t^{\lambda -1}}{(\lambda -1)!}\Theta (t-T').
\end{equation*}
Now, using the displacement theorem for the Laplace transform with $e^{-s(\langle T_R \rangle - T')}$, as well as the fact that $\Theta ((t - (\langle T_R \rangle - T'))-T')=\Theta (t - \langle T_R \rangle)$, we have

\begin{equation*}
    \mathcal{L}^{-1}\left\{\left(\frac{k'}{s + k'}\right)^{\lambda} \frac{\Gamma(\lambda, (s + k')T')}{(\lambda-1)!}e^{-s(\langle T_R \rangle - T')}\right\} = \frac{k'^{\lambda } e^{-k'(t - (\langle T_R \rangle - T'))} (t - (\langle T_R \rangle - T'))^{\lambda -1}}{(\lambda -1)!}\Theta(t - \langle T_R \rangle). 
\end{equation*}
Substituting in $\langle T_R\rangle-T'=\mu_3$ and $k' = k \Lambda$, we can rewrite the above expression as

\begin{equation*}
    f_2(t) \equiv \frac{(k\Lambda)^{\lambda}(t-\mu_3)^{\lambda-1}e^{-k\Lambda(t-\mu_3)}}{(\lambda-1)!}\Theta(t-\langle T_R \rangle).
\end{equation*}
Next, we rewrite $t-\mu_3$ in terms of $t-\langle T_R \rangle$,
\begin{equation*}
    t - \mu_3 
    = t - \langle T_R \rangle + \frac{\langle T_R \rangle}{\Lambda} = \frac{\Lambda(t-\langle T_R \rangle) + \langle T_R \rangle}{\Lambda},
\end{equation*}
which we substitute in to obtain, after simplification, 

\begin{equation*}
    f_2(t) = \Lambda \frac{k^{\lambda}[\Lambda(t-\langle T_R \rangle) + \langle T_R \rangle]^{\lambda-1}e^{-k[\Lambda(t-\langle T_R \rangle) + \langle T_R \rangle]}}{(\lambda-1)!}\Theta(t-\langle T_R \rangle).
\end{equation*}

Defining $x \equiv \Lambda(t-\langle T_R \rangle) + \langle T_R \rangle$ the derivative of which is $\dfrac{dx}{dt} = \Lambda$, we can identify $f_2$ as the density of an $\text{Erlang}(\lambda,k)$ distribution in $x$:
\begin{equation*}
    f_2(t) = \Lambda f_X\big(\Lambda(t-\langle T_R \rangle) + \langle T_R \rangle\big)\Theta(t-\langle T_R \rangle) = \left|\frac{dx}{dt}\right|f_X\big(x(t)\big)\Theta(t-\langle T_R \rangle),
\end{equation*}
where the identity $f_2(t)=|dx/dt|f_X(x(t))$ is the standard change-of-variables formula for probability densities \cite{CasellaBerger2002}. Hence, on $t\ge\langle T_R\rangle$, $f_2$ is the
density of $t=\langle T_R\rangle+(X-\langle T_R\rangle)/\Lambda$ restricted to $X>\langle T_R\rangle$, since that function is increasing and maps
$X=\langle T_R\rangle$ to $t=\langle T_R\rangle$. In sum, we have shown that, as $N\to\infty$, the FPT converges in distribution, by the continuity theorem for Laplace transforms \cite{Feller1966}, to
\begin{equation}
    \boxed{Y =
    \begin{cases}
    X & \text{if} \ X \leq \langle T_R \rangle, \\
    \langle T_R \rangle + \dfrac{X - \langle T_R \rangle}{\Lambda} & \text{if} \ X > \langle T_R \rangle,
    \end{cases}
    \quad\text{with}\ X\sim\mathrm{Erlang}(\lambda,k).}
    \label{eq:Y_ninf}
\end{equation}

Since the theorem also gives convergence of the corresponding moments, $\CV_T^2$ converges to the squared coefficient of variation $\CV_Y^2$ of $Y$.

Next, we proceed to prove that $\CV_Y^2$ admits a unique minimum in $\langle T_R \rangle$ in the limit of deterministic replication.

\subsection{Uniqueness of minimum for deterministic replication.}

Recall that $\CV_Y^2=\langle Y^2\rangle/\langle Y\rangle^2-1$ is the squared coefficient of variation of the FPT for deterministic replication. With $f\equiv f_X$ the density of Erlang$(\lambda,k)$, as before, we also define the following quantities,
\begin{align}
    G(\langle T_R \rangle) &= \int_{\langle T_R \rangle}^{\infty} f(x) dx = \mathbb{P}(X> \langle T_R \rangle), \\
    \aleph(\langle T_R\rangle) &= \int_{0}^{\langle T_R \rangle} x f(x) dx, \\
    \beta(\langle T_R \rangle) &= \int_{0}^{\langle T_R\rangle} x^2 f(x) dx,\quad\text{and}  \\
    E(\langle T_R \rangle) &= \int_{\langle T_R \rangle}^{\infty} G(x) dx = \int_{\langle T_R \rangle}^{\infty} (x -\langle T_R \rangle) f(x) dx.
\end{align}
Next, we express the first and second moments $\langle Y\rangle$ and $\langle Y^2\rangle$, respectively, in terms of these quantities.

Writing $\int_{\langle T_R\rangle}^\infty x f(x) dx$ as $\int_{\langle T_R\rangle}^\infty[\langle T_R\rangle+(x-\langle T_R\rangle)]f(x) dx$, we find
\begin{equation}
    \mu_1 - \aleph(\langle T_R \rangle) = \langle T_R \rangle G(\langle T_R\rangle) + E(\langle T_R \rangle).
    \label{eq:identity_Ninf_1}
\end{equation}
Furthermore, making use of the above definitions, we can write
$\langle Y\rangle=\aleph+\langle T_R\rangle G+E/\Lambda$; similarly, expanding $\big[\langle T_R\rangle(1-1/\Lambda)+X/\Lambda\big]^2$ for $X>\langle T_R\rangle$, we obtain a corresponding expression for $\langle Y^2\rangle$. With Eq.~\eqref{eq:identity_Ninf_1}, the first two moments of $Y$ then read
\begin{align}
    \langle Y \rangle &= \frac{\mu_1}{\Lambda} + \frac{\Lambda-1}{\Lambda}[\aleph(\langle T_R\rangle) + \langle T_R \rangle G(\langle T_R \rangle)]\quad\text{and} \\
    \langle Y^2 \rangle &= \frac{\langle t^2\rangle_{(1)}}{\Lambda^2} + \frac{\Lambda^2 -1}{\Lambda^2}\beta(\langle T_R \rangle) + \left[\frac{(\Lambda-1)\langle T_R\rangle}{\Lambda}\right]^2 G(\langle T_R \rangle) + \frac{2(\Lambda-1)}{\Lambda^2}\langle T_R \rangle (\mu_1 - \aleph(\langle T_R \rangle)).
\end{align}
Differentiating, with $d\aleph/d\langle T_R\rangle=\langle T_R\rangle f(\langle T_R\rangle)$, $d\beta/d\langle T_R\rangle=\langle T_R\rangle^2 f(\langle T_R\rangle)$, and $dG/d\langle T_R\rangle=-f(\langle T_R\rangle)$, we find that the terms that are proportional to $f(\langle T_R\rangle)$ cancel, which gives
\begin{align}
    &\frac{d\langle Y\rangle}{d\langle T_R\rangle} = \frac{\Lambda-1}{\Lambda} G(\langle T_R \rangle) > 0\quad\text{and} \label{eq:dY}\\
    &\frac{d\langle Y^2\rangle}{d\langle T_R\rangle} = 2\frac{\Lambda-1}{\Lambda}\left[\langle T_R \rangle G(\langle T_R\rangle) + \frac{E(\langle T_R \rangle)}{\Lambda}\right] > 0. \label{eq:dY2}
\end{align}
Hence, both moments are strictly increasing in $\langle T_R\rangle$.

\begin{lemma}\label{lem:crit}
For $\langle T_R\rangle>0$,
\begin{equation}
    \frac{d\CV_Y^2}{d\langle T_R\rangle}
    =\frac{2(\Lambda-1)}{\Lambda\langle Y\rangle^{3}}\Psi(\langle T_R\rangle),
    \quad\text{with}\
    \Psi(\langle T_R\rangle)\equiv
    \left(\langle T_R\rangle G+\frac{E}{\Lambda}\right)\langle Y\rangle-\langle Y^2\rangle G.
    \label{eq:dFstar}
\end{equation}
Correspondingly, the critical points of $\CV_Y^2$ are the zeros of $\Psi$, with $\CV_Y^2$ decreasing where $\Psi<0$ and increasing where $\Psi>0$.
\end{lemma}
\begin{proof}
Since $\CV_Y^2=\langle Y^2\rangle/\langle Y\rangle^2-1$,
\begin{equation*}
    \frac{d\CV_Y^2}{d\langle T_R\rangle}
    =\frac{1}{\langle Y\rangle^{3}}\left(\langle Y\rangle\frac{d\langle Y^2\rangle}{d\langle T_R\rangle}
    -2\langle Y^2\rangle\frac{d\langle Y\rangle}{d\langle T_R\rangle}\right).
\end{equation*}
Then, substitution of Eqs.~\eqref{eq:dY} and \eqref{eq:dY2} gives the result.
\end{proof}

\begin{lemma}\label{lem:derivative_psi}
For all $\langle T_R\rangle>0$,
\begin{equation}
    \frac{d\Psi}{d\langle T_R\rangle}
    =\frac{\Lambda-1}{\Lambda}G\aleph
    +f(\langle T_R\rangle)\big[\langle Y^2\rangle-\langle T_R\rangle\langle Y\rangle\big].
\end{equation}
In particular, $d\Psi/d\langle T_R\rangle>0$ at every zero of $\Psi$.
\end{lemma}
\begin{proof}
Since $dG/d\langle T_R\rangle=-f(\langle T_R\rangle)$ and $dE/d\langle T_R\rangle=-G$, $d(\langle T_R\rangle G+E/\Lambda)/d\langle T_R\rangle
=(\Lambda-1)G/\Lambda-\langle T_R\rangle f(\langle T_R\rangle)$. Differentiating $\Psi$ and using Eqs.~\eqref{eq:dY} and \eqref{eq:dY2}, we find
\begin{equation*}
    \frac{d\Psi}{d\langle T_R\rangle}
    =\frac{\Lambda-1}{\Lambda}G\left[\langle Y\rangle-\langle T_R\rangle G-\frac{E}{\Lambda}\right]
    +f\left(\langle T_R\rangle\right)\big[\langle Y^2\rangle-\langle T_R\rangle\langle Y\rangle\big].
\end{equation*}
At zeroes of $\Psi$, $\langle Y^2\rangle=\langle Y\rangle[\langle T_R\rangle+E/(\Lambda G)]$; hence, the second bracketed term above equals $\langle Y\rangle E/(\Lambda G)>0$ then; since the first term equals $\aleph(\langle T_R\rangle)$ and is hence also positive, $d\Psi/d\langle T_R\rangle>0$ there.
\end{proof}

\begin{theorem}[Uniqueness of minimum]\label{thm:unique} For deterministic replication and any $\lambda\ge1$ and $\Lambda>1$, $\CV_Y^2$ has exactly one critical point $\langle T_R\rangle^{*}$ in $(0,\infty)$ which is the global minimum of $\CV_Y^2$, with $\CV_Y^2<1/\lambda$ for all $\langle T_R\rangle>0$.
\end{theorem}
\begin{proof}
We first show the existence of a zero of $\Psi$: as $\langle T_R\rangle\to0^{+}$, $G\to1$, $E\to\mu_1$, $\langle Y\rangle\to\mu_1/\Lambda$, and $\langle Y^2\rangle\to\langle t^2\rangle_{(1)}/\Lambda^2$, which implies $\Psi\to(\mu_1^2-\langle t^2\rangle_{(1)})/\Lambda^2=-\mu_1^2/(\lambda\Lambda^2)<0$. On the other hand, we can write $\Psi=G(\langle T_R\rangle\langle Y\rangle-\langle Y^2\rangle)+E\langle Y\rangle/\Lambda$ for $\langle T_R\rangle>0$. Since $X/\Lambda\le Y\le X$, we have $\langle Y\rangle\ge\mu_1/\Lambda$ and $\langle Y^2\rangle\le\langle t^2\rangle_{(1)}$; hence, both terms in $\Psi$ are positive for $\langle T_R\rangle>\Lambda\langle t^2\rangle_{(1)}/\mu_1$. By continuity, $\Psi$ must have at least one zero in $(0,\infty)$.

Next, we prove that $\Psi$ admits a unique zero. Suppose that $\Psi$ has two zeros $t_1<t_2$. By Lemma~\ref{lem:derivative_psi}, $\Psi>0$ on some interval $(t_1,t_1+\eta)$, so the set $S=\{t\in[t_1+\eta/2,t_2] : \Psi(t)=0\}$ is non-empty and closed; let $t_3=\min S$. Then, $\Psi$ has no zero in $(t_1,t_3)$ and is positive for $t\gtrsim t_1$, i.e.\ for $t$ greater than, but close to, $t_1$, which gives $\Psi>0$ on $(t_1,t_3)$, by continuity. However, $\Psi(t_3)=0$ and $d\Psi/d\langle T_R\rangle\big|_{t=t_3}>0$ imply $\Psi<0$ for $t\lesssim t_3$, a contradiction.

In conclusion, $\Psi$ has exactly one zero $\langle T_R\rangle^{*}$, with $\Psi<0$ for $\langle T_R\rangle<\langle T_R\rangle^{*}$ and $\Psi>0$ for $\langle T_R\rangle>\langle T_R\rangle^{*}$. By Lemma~\ref{lem:crit}, $\CV_Y^2$ decreases on $(0,\langle T_R\rangle^{*})$ and increases on $(\langle T_R\rangle^{*},\infty)$; hence, $\langle T_R\rangle^{*}$ is the only critical point of $\CV_Y^2$ and its global minimum. Since $\CV_Y^2\to1/\lambda$ for both $\langle T_R\rangle\to0^{+}$ and $\langle T_R\rangle\to\infty$ as $Y\to X/\Lambda$ and $Y\to X$ in those limits, respectively, $\CV_Y^2<1/\lambda$ on $(0,\infty)$.
\end{proof}

\begin{corollary}[Absence of peaks]\label{cor:nopeaks}
For deterministic replication, i.e.\ in the limit of $N\to\infty$, the noise has no peak: it lies below the
counting limit for every $\langle T_R\rangle>0$.
\end{corollary}

\section{Location of minimum in the deterministic limit}

Finally, we approximate the location of the unique minimum $\langle T_R\rangle^\ast$ of $\CV_Y^2$, asymptotically as $\lambda\to\infty$, in the deterministic limit ($N\to\infty$).

Recall the definition of the ``clipped" variable $Y$ that is derived from $X \sim \text{Erlang}(\lambda,k)$ as $N\to\infty$, with
\begin{equation}
    Y = X - (1 - \theta)(X - \langle T_R \rangle)^{+},\quad\text{where}\
    \theta \equiv \frac{1}{\Lambda}\ \text{and}\
    (\cdot)^{+} \equiv \max(\cdot,0).
    \label{loc:eq:Ydef}
\end{equation}
Hence, $Y = X$ for $X \leq \langle T_R \rangle$ and $Y = \langle T_R \rangle + \theta(X - \langle T_R \rangle)$ otherwise. Furthermore, $\CV_Y^2 = \operatorname{Var}(Y)/\mathbb{E}[Y]^2$. Throughout, $\Lambda>1$ is fixed. By Theorem~\ref{thm:unique}, the minimum of $\CV_Y^2$ exists and is unique; by Lemma~\ref{lem:crit}, it is characterised by $\Psi=0$, that is,
\begin{equation}
    \left(\langle T_R \rangle G + \frac{E}{\Lambda}\right) \langle Y \rangle = \langle Y^2 \rangle G ,
    \label{loc:eq:known}
\end{equation}
with $G = \mathbb{P}(X > \langle T_R \rangle)$ and $E = \mathbb{E}[(X - \langle T_R \rangle)^{+}]$.

\begin{lemma}[Functional dependence]
\label{lem:red_param}
    The function $\CV_Y^2$ depends on $k$ and $\langle T_R \rangle$ only through the product $x = k\langle T_R \rangle$; hence, $\langle T_R \rangle^{*} = x^{*}(\lambda,\Lambda)/k$.
\end{lemma}

\begin{proof}
    If $X \sim \text{Erlang}(\lambda,k)$, then $k X \sim \text{Erlang}(\lambda,1)$. Multiplying \eqref{loc:eq:Ydef} by $k$ and using the positive homogeneity of $\cdot\mapsto (\cdot)^{+}$, we can write $kY = (k X) - (1-\theta)[(k X) - x]^{+}$, with $x = k\langle T_R \rangle$; in other words, $kY$ is derived from $kX \sim \text{Erlang}(\lambda,1)$, with threshold $x$. Since $\CV_Y^2$ is invariant under multiplication by a positive constant, it depends only on $(\lambda,\Lambda,x)$. 
\end{proof}
We may therefore set $k=1$, corresponding to $x=\langle T_R\rangle$, write $p_{\lambda}(t)=t^{\lambda-1}e^{-t}/\Gamma(\lambda)$ for the density, and abbreviate
\begin{equation}
    G \equiv\mathbb{P}(X > x),\quad \bar{G} \equiv 1 - G, \quad\text{and}\quad g(\lambda,x) \equiv x p_{\lambda}(x) = \frac{x^{\lambda}e^{-x}}{\Gamma(\lambda)}.
    \label{loc:eq:Gg}
\end{equation}

\begin{lemma}[Variance balance]
\label{lem:var_balance}
    The condition in \eqref{loc:eq:known} is equivalent to
    \begin{equation}
     \mathbb{E}[Y] \bar{G}(\mathbb{E}[Y \mid X > x] - \mathbb{E}[Y \mid X \leq x]) = \operatorname{Var}(Y).
    \label{loc:eq:v2}
\end{equation}
\end{lemma}
\begin{proof}
    On $\{X > x\}$, we have $Y = x + \theta(X-x)$ and, hence,
\begin{equation*}
    \mathbb{E}[Y \mathbf{1}\{X>x\}] = x\mathbb{P}(X>x) + \theta  \mathbb{E}[(X-x)\mathbf{1}\{X>x\}] = x G + \frac{E}{\Lambda},
\end{equation*}
which is precisely the bracketed term in \eqref{loc:eq:known}. Hence, \eqref{loc:eq:known} reads $\mathbb{E}[Y\mathbf{1}\{X>x\}]\mathbb{E}[Y] = \mathbb{E}[Y^2] G$. Dividing by $G > 0$ and using $\mathbb{E}[Y\mathbf{1}\{X>x\}] = G \mathbb{E}[Y \mid X>x]$, we find $\mathbb{E}[Y \mid X>x]\mathbb{E}[Y] = \mathbb{E}[Y^2]$; subtraction of $\mathbb{E}[Y]^2$ yields $\mathbb{E}[Y](\mathbb{E}[Y\mid X>x] - \mathbb{E}[Y]) = \operatorname{Var}(Y)$.
Finally, the tower property $\mathbb{E}[Y] = G \mathbb{E}[Y\mid X>x] + \bar{G}\mathbb{E}[Y \mid X \leq x]$ gives $\mathbb{E}[Y\mid X>x] - \mathbb{E}[Y] = \bar{G}(\mathbb{E}[Y\mid X>x] - \mathbb{E}[Y\mid X\leq x])$, which is \eqref{loc:eq:v2}. 
\end{proof}
For future reference, we note that the proof of Lemma~\ref{lem:var_balance} implies
\begin{equation}
\Psi(\langle T_R\rangle)= G\big\{\mathbb{E}[Y] \bar{G}(\mathbb{E}[Y \mid X > x] - \mathbb{E}[Y \mid X \leq x]) - \operatorname{Var}(Y)\big\};
\label{loc:eq:psi}
\end{equation}
recall Eq.~\eqref{eq:dFstar}.

\begin{lemma}[Mean gap]\label{lem:mean_gap}
    For every $x > 0$,
\begin{equation}
    \mathbb{E}[X \mid X > x] = \lambda + \frac{g}{G}
    \quad\text{and}\quad
    \mathbb{E}[X \mid X \leq x] = \lambda - \frac{g}{\bar{G}}.
    \label{loc:eq:condmeans}
\end{equation}
\end{lemma}
\begin{proof}
   Since $t p_{\lambda}(t)=\lambda p_{\lambda+1}(t)$, integration over $(x,\infty)$ gives $\mathbb{E}[X\mathbf{1}\{X>x\}]=\lambda\mathcal{Q}_{\lambda+1}(x)$, where $\mathcal{Q}_\nu(x)=\Gamma(\nu,x)/\Gamma(\nu)$ is the upper regularised incomplete Gamma function $\mathcal{Q}_\lambda(\mu)\equiv\sum_{n<\lambda}e^{-\mu}\mu^{n}/n!$ \cite{NIST:DLMF}, whence $G=\mathcal{Q}_\lambda(x)$. Integration by parts gives the recurrence $\mathcal{Q}_{\lambda+1}(x)=\mathcal{Q}_{\lambda}(x)+g/\lambda$, which implies $\mathbb{E}[X\mathbf{1}\{X>x\}]=\lambda G+g$.
    
    Division by $G$ gives the first identity. Subtraction from $\mathbb{E}[X] = \lambda$ yields $\mathbb{E}[X\mathbf{1}\{X\leq x\}] = \lambda\bar{G} - g$; division by $\bar{G}$ then gives the second identity.
\end{proof}

Since $Y = X$ below the threshold $x$ and $Y = \theta X + (1-\theta)x$ above it, Lemma~\ref{lem:mean_gap} specifies the gap in \eqref{loc:eq:v2}: writing $x = \lambda - \delta$, we have
\begin{equation}
    \mathbb{E}\left[Y \mid X>x\right] - \mathbb{E}\left[Y \mid X \leq x\right]
    = \theta\left(\lambda + \frac{g}{G}\right) + \left(1-\theta\right)x - \left(\lambda - \frac{g}{\bar{G}}\right)
    = -\left(1-\theta\right)\delta + \theta\frac{g}{G} + \frac{g}{\bar{G}}.
    \label{loc:eq:gap}
\end{equation}

\subsection{Derivation of balance relation}

Let $x=\lambda-\delta$ and $z\equiv\delta/\sqrt{\lambda}$. We consider the asymptotic regime of $\lambda\to\infty$ with $z\to\infty$ and $z=o(\lambda^{1/6})$, i.e.
\begin{equation}  
\varepsilon\equiv\frac{z^3}{\sqrt{\lambda}}=\frac{\delta^3}{\lambda^2}\to0 .
\label{loc:eq:eps}
\end{equation}
We note that the quantities $z/\sqrt{\lambda} = \delta/\lambda$, $1/\sqrt{\lambda}$, and $1/\delta$ are all $O(\varepsilon)$ for $z\geq 1$. (Since $\Lambda$ is fixed, the constants implied by $O(\cdot)$ may depend on $\Lambda$.)

\begin{lemma}[Mills ratio]\label{lem:mills_ratio}
    In the above regime,
\begin{equation}
    \frac{\delta\bar{G}}{g} = 1 - z^{-2} + O(z^{-4},\varepsilon).
    \label{loc:eq:mills}
\end{equation}
\end{lemma}
\begin{proof}
    With $x=\lambda-\delta$ and $g=x p_\lambda(x)$,
\begin{equation*}
    \frac{\delta\bar{G}}{g}
    =\frac{\delta}{x}\int_0^{x}\frac{p_\lambda(x-u)}{p_\lambda(x)} du
    =\frac{\delta}{x}\int_0^{x}\exp\left[(\lambda-1)\ln\left(1-\frac{u}{x}\right)+u\right]du.
\end{equation*}
Expanding the logarithm about $u=0$, we find that the exponent equals $-au-bu^2/2+O(\lambda u^3/x^3)$, with $a=(\delta-1)/x$ and $b=(\lambda-1)/x^2$. Since $\ln(1-y)\le-y-y^2/2$, the integrand is bounded by $e^{-au-bu^2/2}$, which implies that the integral is dominated by $u=O(1/a)=O(\sqrt{\lambda}/z)$, where the cubic remainder is $O\big[(z^3\sqrt{\lambda})^{-1}\big]=O(\varepsilon)$; moreover, 
extending the upper limit of integration to infinity, we only induce an exponentially small change in the integral. Using $\int_0^\infty e^{-au-bu^2/2} du=\Phi(-w)/\bigl(\sqrt{b} \varphi(w)\bigr)$ with $w=a/\sqrt{b}$, where $\varphi$ and $\Phi$ are the standard normal density and distribution function, respectively, we obtain
\begin{equation*}
    \frac{\delta\bar{G}}{g}
    =\frac{\delta}{\sqrt{\lambda-1}} \frac{\Phi(-w)}{\varphi(w)}[1+O(\varepsilon)],
    \quad\text{with}\ w=\frac{\delta-1}{\sqrt{\lambda-1}} .
\end{equation*}
Since both $\delta/\sqrt{\lambda-1}$ and $w$ equal $z[1+O(\varepsilon)]$, and since the logarithmic derivative of $w\mapsto w\Phi(-w)/\varphi(w)$ is bounded for $w\ge1$, the Mean Value Theorem implies $\delta\bar{G}/g=z\Phi(-z)/\varphi(z)[1+O(\varepsilon)]$. The Mills expansion $z\Phi(-z)/\varphi(z)=1-z^{-2}+O(z^{-4})$ then gives the claim.
\end{proof}

\begin{proposition}\label{prop:balance}
    In the above regime, with $g(\lambda,x)=O(1)$, Eq.~\eqref{loc:eq:psi} reduces to
\begin{equation}
    \boxed{\ \Psi = G\lambda\theta^2\Big\{\frac{g}{\theta}\big[1+(\Lambda-1)z^{-2}+O(z^{-4},\varepsilon)\big]-1+O(\varepsilon)\Big\} \ }
    \label{loc:eq:balance3}
\end{equation}
In particular,
\begin{equation}
    g(\lambda,x^{*}) = \frac{1}{\Lambda}\frac{1}{1 + (\Lambda-1)z^{-2}}\big[1 + O(z^{-4},\varepsilon)\big]
    \label{loc:eq:balance2}
\end{equation}
and, hence,
\begin{equation}
g(\lambda,x^{*}) = \frac{1}{\Lambda}\big[1 + O(z^{-2},\varepsilon)\big]
\label{loc:eq:balance}
\end{equation}
at zeroes $x^\ast$ of $\Psi$.
\end{proposition}

\begin{proof}
    We rewrite the three factors in \eqref{loc:eq:psi} in terms of $\theta$, $z$, and $\lambda$. By Lemma~\ref{lem:mills_ratio} and our assumption that $g=O(1)$, we have $\bar{G}=O(g/\delta)=O(\varepsilon)$.

   By \eqref{loc:eq:Ydef}, $\mathbb{E}[Y] = \lambda - (1-\theta)\mathbb{E}[(X-x)^{+}]$ and $\mathbb{E}[(X-x)^{+}] \leq \mathbb{E}[|X-x|] \leq \delta + \sqrt{\lambda}$, so $\mathbb{E}[Y] = \lambda[1 + O(\varepsilon)]$.

   Next, multiplying the exact gap in \eqref{loc:eq:gap} by $\bar{G}$, we find
    \begin{equation}
        \bar{G}(\mathbb{E}[Y \mid X>x] - \mathbb{E}[Y \mid X \leq x])
        = g - (1-\theta)\delta\bar{G} + \theta g\frac{\bar{G}}{G}.
    \label{loc:eq:gapexact}
\end{equation}
By Lemma \ref{lem:mills_ratio}, $(1-\theta)\delta\bar{G} = (1-\theta)g[1 - z^{-2} + O(z^{-4},\varepsilon)]$. Furthermore, $\theta g\bar{G}/G = O(g\varepsilon)$ due to $\bar{G} = (g/\delta)[1+o(1)]$; since $g=O(1)$, by assumption, $\bar{G}=O(\delta^{-1})=O(\varepsilon)$, while $G=O(1)$. Using $1-\theta = \theta(\Lambda-1)$, we can write
\begin{align*}
    \bar{G}(\mathbb{E}[Y \mid X>x] - \mathbb{E}[Y \mid X \leq x])
    &= g- (1-\theta) g\big[1-z^{-2} + O(z^{-4}, \varepsilon)\big] +  O(g\varepsilon) \\
    &= g\big[\theta + (1-\theta) z^{-2} + O(z^{-4},\varepsilon)\big] \\
    &= \theta g\big[1 + (\Lambda-1)z^{-2} + O(z^{-4},\varepsilon)\big],
\end{align*}
where we have used $(1-\theta)/\theta = \Lambda -1$.

Finally, since the Erlang density is log-concave, the right-truncated variance $\operatorname{Var}(X\mid X\le x)$ is non-decreasing in $x$~\cite[Sec.~2 and Prop.~4]{Burdett1996}; hence, $\operatorname{Var}(X \mid X \leq x)\le\operatorname{Var}(X)=\lambda$. The law of total variance for $X$, together with Lemma~\ref{lem:mean_gap} and Lemma~\ref{lem:mills_ratio} -- which imply that the gap between the conditional means of $X$ is $O\left(\delta\right)$ -- then gives $\operatorname{Var}(X \mid X > x) = \lambda[1 + O(\varepsilon)]$. On $\{X>x\}$, the map $X \mapsto Y$ is affine with slope $\theta$, so the law of total variance for $Y$ reads
\begin{equation*}
    \operatorname{Var}(Y) = \theta^2 G\operatorname{Var}(X \mid X>x) + \bar{G}\operatorname{Var}(X \mid X\leq x) + O(\bar{G}\delta^2)
    = \theta^2\lambda[1 + O(\varepsilon)],
\end{equation*}
since $\bar{G} = O(\varepsilon)$ and $\bar{G}\delta^2 = O(g\delta) = O(\lambda\varepsilon)$.

Substituting the estimates for the three factors into \eqref{loc:eq:psi}, we conclude that

\begin{align*}
\Psi = G\lambda\theta g\big[1 + (\Lambda-1)z^{-2} + O(z^{-4},\varepsilon)\big]-G\lambda\theta^2[1 + O(\varepsilon)],
\end{align*}
which is precisely \eqref{loc:eq:balance3}. At zeroes $x^\ast$ of $\Psi$, we can solve the expression in curly brackets for $g$ to find \eqref{loc:eq:balance2} and, hence, \eqref{loc:eq:balance}.
\end{proof}

\subsection{Solution of balance relation}

Having derived a condition that needs to be satisfied by the probability density $g$ at the unique minimum of $\CV_Y^2$, we now estimate the corresponding value of $x^\ast$.

\begin{lemma}[Gaussian decay]
\label{lem:gauss_decay}
    The function $g(\lambda,x)$ is strictly increasing in $x$ on $(0,\lambda)$ and maximised at $x = \lambda$, with $g_{\max} = \sqrt{\lambda/2\pi}[1+O(\lambda^{-1})]$. Moreover, for $\delta = o(\lambda^{2/3})$ and $z\geq 1$,
    \begin{equation}
    \ln g(\lambda,\lambda-\delta) = \frac{1}{2}\ln\frac{\lambda}{2\pi} - \frac{z^2}{2} + O(\varepsilon).
    \label{loc:eq:decay}
\end{equation}
\end{lemma}
\begin{proof}
    Since the derivative $d[\lambda \ln x - x]/dx = \lambda/x - 1$ vanishes only at $x = \lambda$, it follows that $g(\lambda,x)=x^\lambda e^{-x}/\Gamma(\lambda)$ is unimodal, with its maximum there; the estimate for $g_{\max}$ is obtained from Stirling's formula: $\Gamma(\lambda) = \sqrt{2\pi}\lambda^{\lambda-1/2}e^{-\lambda}[1+O(\lambda^{-1})]$. For the decay estimate, Stirling's formula gives $g(\lambda,\lambda-\delta) = \sqrt{\lambda/2\pi}(1-\delta/\lambda)^{\lambda}e^{\delta}[1+O(\lambda^{-1})]$. Then, taking the logarithm and expanding, we find
\begin{equation}
    \lambda\ln\Big(1-\frac{\delta}{\lambda}\Big) + \delta = -\frac{\delta^2}{2\lambda} - \frac{\delta^3}{6\lambda^2} - \cdots ,
    \label{loc:eq:skew}
\end{equation}

where $\delta^2/(2\lambda) = z^2/2$ and the cubic (skewness) term evaluates to $\delta^3/\lambda^2 = \varepsilon$, which is $o(1)$ exactly when $\delta = o(\lambda^{2/3})$; an analogous argument shows that the tail in \eqref{loc:eq:skew} is negligible. Finally, $\ln(1+O(\lambda^{-1}))=O(\lambda^{-1})=O(\varepsilon)$, as $\varepsilon = z^3/\sqrt{\lambda} \geq \lambda{-1/2} \geq \lambda^{-1}$ for $z\geq 1$.
\end{proof}

Equation \eqref{loc:eq:skew} shows that $\varepsilon$ measures the deviation of the function $p_\lambda$ from a Gaussian, to leading order, which motivates the asymptotic regime introduced beforehand.

Next, we show that the unique minimum $x^\ast$ of $\CV_Y^2$ lies in the asymptotic regime specified above.
\begin{lemma}[Localisation]
\label{lem:local}
    Define $L\equiv\ln(\lambda\Lambda^2/(2\pi))$, and let
    \begin{equation*}
        z_\pm\equiv\sqrt{L \pm 1}\quad\text{and}\quad x_\pm=\lambda - \sqrt{\lambda}z_\pm.
    \end{equation*}
    Then, $0<x_+<x_-<\lambda$ and $\Psi(x_+)<0<\Psi(x_-)$, i.e.\ the unique minimiser $x^\ast$ of $\CV_Y^2$ satisfies $x_+<x^\ast<x_-$. Consequently, $L-1 < (z^\ast)^2 < L + 1$ for $z^\ast=(\lambda - x^\ast)/\sqrt{\lambda}$, which implies that $x^\ast$ lies in the above regime, with $g(\lambda,x^\ast)=O(1)$. 
\end{lemma}
\begin{proof}
    Since $L=\ln\lambda+O(1)$, we have $z_\pm\to\infty$, with $\varepsilon_\pm\equiv z_\pm^3\sqrt{\lambda}=(L \pm 1)^{3/2}/\sqrt{\lambda}\to 0$; hence, $x_\pm$ lie in the above regime, and $0 < x_+ < x_- < \lambda$ for large $\lambda$.
    
    By Lemma~\ref{lem:gauss_decay} and $\frac{1}{2}\ln(\lambda/(2\pi))-\frac{1}{2}L=-\ln\Lambda$,
    \begin{equation*}
      \ln g(\lambda,x_\pm) = -\ln\Lambda \mp \frac{1}{2} + O(\varepsilon_\pm),
    \end{equation*}
    that is, $g(\lambda,x_\pm)=\theta e^{\mp 1/2}[1 + o(1)]=O(1)$; Proposition~\ref{prop:balance} hence applies at $x_\pm$. Since $z_\pm^{-2}=O(L^{-1})\to 0$ as $\lambda\to\infty$, \eqref{loc:eq:balance3} yields $\Psi(x_\pm) = G\lambda\theta^2\big[e^{\mp 1/2} - 1 + o(1)\big]$. As $G\lambda\theta^2>0$, it follows that $\Psi(x_+)<0<\Psi(x_-)$ for $\lambda$ sufficiently large.
    
    Since $\Psi$ is continuous, the Intermediate Value Theorem implies that the unique zero $x^\ast$ of $\Psi$ -- and, hence, the unique global minimiser of $\CV_Y^2$ -- must satisfy $x_+<x^\ast<x_-$; recall Theorem~\ref{thm:unique}. Equivalently, $z_-<z^\ast<z_+$, as claimed; in particular, $z^\ast\to\infty$ and $\varepsilon^\ast\equiv (z^\ast)^3/\sqrt{\lambda}=O\big(L^{3/2}/\sqrt{\lambda}\big)\to 0$ as $\lambda\to\infty$. Finally, since $g(\lambda,x)$ is increasing in $x$ on $(0,\lambda)$, by Lemma~\ref{lem:gauss_decay}, $g(\lambda,x_+)<g(\lambda,x^\ast)<g(\lambda,x_-)$, which implies $g(\lambda,x^\ast)=O(1)$. 
\end{proof}

Finally, we combine the results of Proposition~\ref{prop:balance} and Lemma~\ref{lem:gauss_decay}:
\begin{theorem}\label{thm:optimal}
     As $\lambda \to \infty$ with $\Lambda > 1$ fixed,
\begin{equation}
    \boxed{\ \langle T_R \rangle^{*} = \frac{1}{k}\left\{\lambda - \sqrt{\lambda \ln\frac{\lambda\Lambda^2}{2\pi}}\left[1 + O[(\ln\lambda)^{-2}]\right]\right\}. \ }
    \label{loc:eq:closed}
\end{equation}
\end{theorem}
\begin{proof}
By Lemma~\ref{lem:local}, $x^\ast$ satisfies the conditions of Proposition~\ref{prop:balance}; in particular, $g(\lambda,x^\ast)=O(1)$. By \eqref{loc:eq:balance}, $\ln g(\lambda,x^\ast)=-\ln\Lambda+O\big[(z^\ast)^{-2},\varepsilon\big]$, whereas by Lemma~\ref{lem:gauss_decay}, $\ln g(\lambda,\lambda-\delta^\ast)=\frac{1}{2}\ln(\lambda/2\pi)-(z^\ast)^2/2+O(\varepsilon^\ast)$. Equating the two,
\begin{equation}
    (z^\ast)^2 = L + O\big[(z^\ast)^{-2},\varepsilon^\ast\big],
    \label{loc:eq:s2}
\end{equation}
where $L\equiv\ln(\lambda\Lambda^2/2\pi)$, as before. Lemma~\ref{lem:local} gives $(z^\ast)^2 = L + O(1)$, which implies $O\big[(z^\ast)^{-2}\big] = O(L^{-1})$ and $\varepsilon^\ast = O(L^{3/2}/\sqrt{\lambda}) = o(L^{-1})$. Hence, $z^2 = L[1 + O(L^{-2})]$ and
\begin{equation*}
    \delta^{*} = \sqrt{\lambda}z^\ast = \sqrt{\lambda L}\big[1 + O(L^{-2})\big].
\end{equation*}
Then, $x^{*} = \lambda - \delta^{*}$ and $\langle T_R \rangle^{*} = x^{*}/k$ by Lemma~\ref{lem:red_param}, which gives \eqref{loc:eq:closed}.
\end{proof}

\begin{remark}
    \begin{itemize}
        \item The error term in Eq.~\eqref{loc:eq:closed} multiplies only the correction $\sqrt{\lambda L}$, i.e.\ it is stated for the gap $\delta^{*} = \lambda - k\langle T_R \rangle^{*}$, rather than for $\langle T_R \rangle^{*}$ itself. Equivalently, one could write $k\langle T_R\rangle^\ast=\lambda-\sqrt{\lambda L}+O\big(\sqrt\lambda L^{-3/2}\big)$, which implies a relative error of $O\big[\lambda^{-1/2}(\ln\lambda)^{-3/2}\big]$ in $\langle T_R \rangle^{*}$.
        \item The leading-order asymptotics of $\langle T_R\rangle^\ast$ already follows from Lemma~\ref{lem:local}, which gives 
        \begin{equation*}
            \sqrt{\lambda(L-1)}<\lambda - k\langle T_R\rangle^\ast<\sqrt{\lambda(L+1)}
        \end{equation*}
        for sufficiently large $\lambda$ and, hence, $k\langle T_R\rangle^\ast=\lambda-\sqrt{\lambda L}+O\big(\sqrt{\lambda/L}\big)$. The proof of Theorem~\ref{thm:optimal} combines Proposition~\ref{prop:balance} and Lemma~\ref{lem:gauss_decay} to refine the error in that asymptotics by a factor of $L^{-1}$.
        \item The constant implied by $O[(\ln\lambda)^{-2}]$ in \eqref{loc:eq:closed} is given by $\Lambda - 1$. Indeed, repeating the proof with \eqref{loc:eq:balance2} instead of \eqref{loc:eq:balance}, we find $z^2 = L + 2\ln\big(1 + (\Lambda-1)z^{-2}\big) + O(z^{-4},\varepsilon)= L + 2(\Lambda-1)/L + O(L^{-2}) + O(\varepsilon)$ and, therefore,
        \begin{equation}
            \delta^{*} = \sqrt{\lambda L}\left[1 + \frac{\Lambda-1}{L^2} + o(L^{-2})\right],
             \label{loc:eq:refined}
        \end{equation}
        up to $O(\varepsilon)$-terms.
    \end{itemize}

\end{remark}

\section{Bursty gene expression considering DNA replication}

\begin{figure}[htb]
	\noindent    \includegraphics[width=\textwidth,height=0.855\textheight,keepaspectratio]{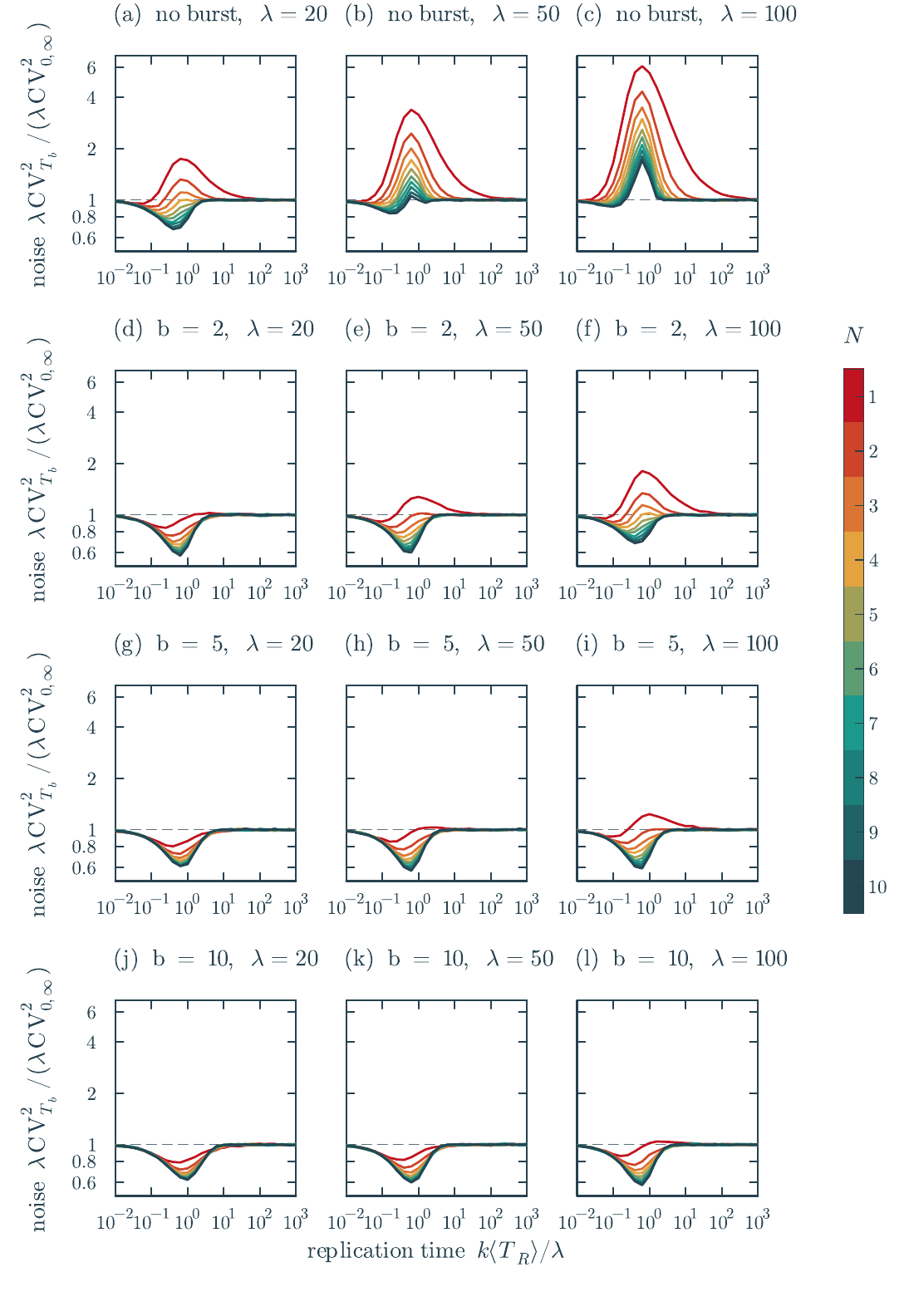}
    \captionsetup{skip=1pt, font=footnotesize}
	\caption{\justifying\textbf{Noise in the first passage time $T_b$ with threshold $\lambda$, as a function of the mean DNA replication time.} Data obtained with the stochastic simulation algorithm (SSA), with $8 \times 10^4$ realisations per point at 26 logarithmically spaced values of $k\langle T_R\rangle/\lambda$. Standard errors, estimated from 50 independent batches of 1600 trajectories, are below $0.8\%$ of $\lambda\CV_{T_b}^2$ at all points (median $0.5\%$), i.e.\ smaller than the line width.
    Plots are shown on a $\log_{10}$-scale in both axes. Rows~2 through 4 correspond to different combinations of burst size $b$ and threshold $\lambda$. Row~1 (no bursting) agrees with the exact results shown in Figure~2(i) of the main text. The qualitative behaviour closely matches the one reported there; however, peaks are seen to disappear with increasing $b$ for $\lambda$ fixed. The burst sizes shown are representative; a similar behaviour is expected for larger values of $b$. Note that, when bursting is present, the common limit $\CV_{0,\infty}^2$ of $\CV_{T_b}^2$ as $\langle T_R \rangle \to 0^+$ and $\langle T_R \rangle \to \infty$ derived above no longer equals $1/\lambda$; the noise is therefore normalised by $\CV_{0,\infty}^2$.}
	\label{fig:bursts}
\end{figure}

Gene expression is typically bursty in both prokaryotes and eukaryotes. To account for this burstiness and to test one of the limitations of our model, we performed stochastic simulations using the Gillespie algorithm~\cite{Gillespie1976, Gillespie2007}, as implemented in the Catalyst package~\cite{Loman2023} for the Julia programming language; see Figure~\ref{fig:bursts}. To include bursty gene expression, we modified the reaction network in \eqref{eq:react_network} so that the mean protein production rate remains constant as the burst size varies, which ensures comparability between simulations. The resulting modified reaction network reads
\begin{equation}
\begin{aligned}
G_0 &\xrightarrow{\alpha} G_1 \xrightarrow{\alpha} G_2 \xrightarrow{\alpha} \cdots \xrightarrow{\alpha} G_N, \\
G_j &\xrightarrow{k/b} G_j + nP \quad\text{for } j \in \{0, \ldots, N-1\}, \\
G_N &\xrightarrow{\Lambda k/b} G_N + nP,
\end{aligned}
\label{eq:react_network_2}
\end{equation}
where $b$ is the mean burst size and the burst size $n$ follows a geometric distribution on $\{0, 1, 2, \ldots\}$, with $P(n) = [1/(1+b)][b/(1+b)]^n$, as described in~\cite{Shahrezaei2008}. Since bursts occur at rate $k/b$ (or $\Lambda k/b$) and $\langle n \rangle = b$, the mean production rate is $k$ (or $\Lambda k$), independently of $b$.

When gene expression is bursty, the limiting values of $\CV_T^2$ for $\langle T_R \rangle \to 0^+$ and $\langle T_R \rangle \to \infty$ no longer equal $1/\lambda$. To derive the correct limit, recall first that $\CV_T^2$ does not depend on the production rate $k$ -- and, by extension, on $\Lambda k$; hence, both limits coincide, and it suffices to consider production at rate $k$. Bursts, including null bursts ($n = 0$), occur at rate $k/b$, with each burst being non-null with probability $b/(1+b)$. By the thinning property of Poisson processes, non-null bursts therefore occur as a Poisson process with rate
\begin{equation*}
\nu = \frac{k}{b} \frac{b}{1+b} = \frac{k}{1+b}.
\end{equation*}
Moreover, due to the memorylessness of the geometric distribution, the size $m$ of a non-null burst is geometrically distributed on $\{1, 2, \ldots\}$, with $P(m) = [1/(1+b)][b/(1+b)]^{m-1}$.

Let $B$ denote the number of non-null bursts needed to reach the threshold $\lambda$. Each of these bursts is preceded by an exponentially distributed waiting time, so that the first-passage time (FPT) is the sum of $B$ independent waiting times:
\begin{equation*}
T_b = \sum_{i=1}^{B} \rho_i, \quad\text{with } \rho_i \sim \text{Exp}(\nu).
\end{equation*}

Since $B$ is independent of the waiting times $\rho_i$, the law of total expectation gives $\langle T_b \rangle = \langle B \rangle/\nu$, while the law of total variance gives

\begin{equation*}
\text{Var}(T_b) = \langle \text{Var}(T_b \mid B) \rangle + \text{Var}(\langle T_b \mid B \rangle) = \frac{\langle B \rangle}{\nu^2} + \frac{\text{Var}(B)}{\nu^2}.
\end{equation*}
Consequently,
\begin{equation*}
\CV_{T_b}^2 = \frac{\langle B \rangle + \text{Var}(B)}{\langle B \rangle^2},
\end{equation*}

To determine the distribution of $B$, consider an integer threshold $\lambda \geq 1$. The first protein molecule is always produced by the first non-null burst. For each subsequent molecule $j = 2, \ldots, \lambda$, two scenarios are possible: either the burst that produced the $(j-1)$-th molecule is still active, which happens with probability $b/(1+b)$, and also produces the $j$-th molecule; or that burst has ended, which occurs with probability $1/(1+b)$, and a new burst is required to produce the $j$-th molecule. Hence,

\begin{equation*}
B = 1 + \text{Binom}\left(\lambda - 1, \frac{1}{1+b}\right),
\end{equation*}
with
\begin{equation*}
\langle B \rangle = \frac{\lambda + b}{1+b}\quad\text{and}\quad\text{Var}(B) = \frac{(\lambda - 1)b}{(1+b)^2}.
\end{equation*}
Substitution of these expressions into $\CV_{T_b}^2$ yields
\begin{equation}
  \lim_{\langle T_R \rangle \to 0^+} \CV^2_{T_b} = \frac{\lambda(1+2b) + b^2}{(\lambda + b)^2},
\label{eq:cv_lim_bursty}
\end{equation}
which recovers $\lim_{\langle T_R \rangle \to 0^+} \CV^2_{T_b} = 1/\lambda$ in the non-bursty limit of $b \to 0$. Note that $\lim_{\langle T_R \rangle \to \infty } \CV^2_{T_b} = \lim_{\langle T_R \rangle \to 0^+} \CV^2_{T_b}\equiv \CV_{0,\infty}^2$. In addition, the mean FPT is
\begin{equation*}
\langle T_b \rangle = \frac{\langle B \rangle}{\nu} = \frac{\lambda + b}{k} = \frac{\lambda}{k} + \frac{b}{k}.
\end{equation*}
The first term is the time required to produce $\lambda$ molecules at the mean rate $k$, as in the non-bursty case, whereas the second term arises from the overshoot of the last burst, which exceeds the threshold by $b$ molecules on average. The same derivation applies to production at rate $\Lambda k$, which changes $\langle T_b \rangle$, but leaves the extreme limits in $\langle T_R \rangle$ of $\CV^2_{T_b}$ unchanged. 

Eq.~\eqref{eq:cv_lim_bursty} allows us to normalise the results shown in Figure~\ref{fig:bursts}, ensuring comparability between plots, and to better understand how the limiting behaviour changes under bursty expression, given that the limits of $\langle T_R \rangle \to 0^+$ and $\langle T_R \rangle \to \infty$ are used extensively throughout this work. In our simulations, we used the same parameters as in Figure~2 of the main text, varying only the mean burst size $b$. Although actual burst sizes may exceed those considered here, the observed trend is clear, and we expect it to persist for larger values of $b$.

\clearpage

\bibliographystyle{apsrev4-2}
\bibliography{article_refs}

@article{sun2019nonlinear,
  title={The nonlinear dynamics and fluctuations of mRNA levels in cell cycle coupled transcription},
  author={Sun, Qiwen and Jiao, Feng and Lin, Genghong and Yu, Jianshe and Tang, Moxun},
  journal={PLoS Comput. Biol.},
  volume={15},
  number={4},
  pages={e1007017},
  year={2019},
  publisher={Public Library of Science, San Francisco, CA USA}
}

@article{peterson2015effects,
  title={Effects of DNA replication on mRNA noise},
  author={Peterson, Joseph R and Cole, John A and Fei, Jingyi and Ha, Taekjip and Luthey-Schulten, Zaida A},
  journal={Proc. Natl. Acad. Sci. U.S.A.},
  volume={112},
  number={52},
  pages={15886--15891},
  year={2015},
  publisher={National Academy of Sciences}
}

@article{dessalles2020models,
  title={Models of protein production along the cell cycle: An investigation of possible sources of noise},
  author={Dessalles, Renaud and Fromion, Vincent and Robert, Philippe},
  journal={PLoS ONE},
  volume={15},
  number={1},
  pages={e0226016},
  year={2020},
  publisher={Public Library of Science, San Francisco, CA USA}
}

@book{Alon2019,
	title = {An Introduction to Systems Biology: Design Principles of Biological Circuits},
	ISBN = {9780429283321},
	url = {http://dx.doi.org/10.1201/9780429283321},
	DOI = {10.1201/9780429283321},
	publisher = {Chapman and Hall/CRC},
	author = {Alon, Uri},
	year = {2019}
}

@ARTICLE{Lappin2006-ao,
	title    = "{HOX} genes: seductive science, mysterious mechanisms",
	author   = "Lappin, Terence R J and Grier, David G and Thompson, Alexander
	and Halliday, Henry L",
	journal  = "Ulster Med. J.",
	volume   =  "75",
	number   =  "1",
	pages    = "23-31",
	year     = "2006",
	language = "en"
}

@article{Montavon2014,
	title = {Hox gene regulation and timing in embryogenesis},
	volume = {34},
	ISSN = {1084-9521},
	url = {http://dx.doi.org/10.1016/j.semcdb.2014.06.005},
	DOI = {10.1016/j.semcdb.2014.06.005},
	journal = {Semin. Cell Dev. Biol.},
	publisher = {Elsevier BV},
	author = {Montavon,  Thomas and Soshnikova,  Natalia},
	year = {2014},
	month = oct,
	pages = {76–84}
}

@article{Young2009,
  title = {Cdx and Hox Genes Differentially Regulate Posterior Axial Growth in Mammalian Embryos},
  volume = {17},
  ISSN = {1534-5807},
  url = {http://dx.doi.org/10.1016/j.devcel.2009.08.010},
  DOI = {10.1016/j.devcel.2009.08.010},
  number = {4},
  journal = {Dev. Cell},
  publisher = {Elsevier BV},
  author = {Young,  Teddy and Rowland,  Jennifer Elizabeth and van de Ven,  Cesca and Bialecka,  Monika and Novoa,  Ana and Carapuco,  Marta and van Nes,  Johan and de Graaff,  Wim and Duluc,  Isabelle and Freund,  Jean-Noël and Beck,  Felix and Mallo,  Moises and Deschamps,  Jacqueline},
  year = {2009},
  month = Oct,
  pages = {516–526}
}

@article{Matthews2022,
	title = {Cell cycle control in cancer},
	volume = {23},
	ISSN = {1471-0080},
	url = {http://dx.doi.org/10.1038/s41580-021-00404-3},
	DOI = {10.1038/s41580-021-00404-3},
	number = {1},
	journal = {Nat. Rev. Mol. Cell Biol.},
	publisher = {Springer Science and Business Media LLC},
	author = {Matthews,  Helen K. and Bertoli,  Cosetta and de Bruin,  Robertus A. M.},
	year = {2022},
	pages = {74–88}
}

@article{Ghusinga2017,
	title = {First-passage time approach to controlling noise in the timing of intracellular events},
	volume = {114},
	ISSN = {1091-6490},
	url = {http://dx.doi.org/10.1073/pnas.1609012114},
	DOI = {10.1073/pnas.1609012114},
	number = {4},
	journal = {Proc. Natl. Acad. Sci. U.S.A.},
	publisher = {Proc. Natl. Acad. Sci. U.S.A.},
	author = {Ghusinga,  Khem Raj and Dennehy,  John J. and Singh,  Abhyudai},
	year = {2017},
	month = Jan,
	pages = {693–698}
}

@article{Zopf2013,
  title = {Cell-Cycle Dependence of Transcription Dominates Noise in Gene Expression},
  volume = {9},
  ISSN = {1553-7358},
  url = {http://dx.doi.org/10.1371/journal.pcbi.1003161},
  DOI = {10.1371/journal.pcbi.1003161},
  number = {7},
  journal = {PLoS Comput. Biol.},
  publisher = {Public Library of Science (PLoS)},
  author = {Zopf,  C. J. and Quinn,  Katie and Zeidman,  Joshua and Maheshri,  Narendra},
  editor = {Kondev,  Jane},
  year = {2013},
  pages = {e1003161}
}

@article{Beentjes2020,
	author  = {Beentjes, C. H. and Perez-Carrasco, R. and Grima, R.},
	title   = {Exact solution of stochastic gene expression models with bursting, cell cycle and replication dynamics},
	journal = {Phys. Rev. E},
	volume  = {101},
	number  = {3},
	pages   = {032403},
	year    = {2020},
	doi     = {10.1103/PhysRevE.101.032403}
}

@article{Cao2020,
	author  = {Cao, Zhixing and Grima, Ramon},
	title   = {Analytical distributions for detailed models of stochastic gene expression in eukaryotic cells},
	journal = {Proc. Natl. Acad. Sci. U.S.A.},
	volume  = {117},
	number  = {9},
	pages   = {4682--4692},
	year    = {2020},
	doi     = {10.1073/pnas.1910888117}
}

@article{JiaSinghGrima2022,
	author  = {Jia, Chen and Singh, Abhyudai and Grima, Ramon},
	title   = {Concentration fluctuations in growing and dividing cells: Insights into the emergence of concentration homeostasis},
	journal = {PLoS Comput. Biol.},
	volume  = {18},
	number  = {10},
	pages   = {e1010574},
	year    = {2022},
	doi     = {10.1371/journal.pcbi.1010574}
}

@article{Yates2017,
	author  = {Yates, Christian A. and Ford, Matthew J. and Mort, Richard L.},
	title   = {A Multi-stage Representation of Cell Proliferation as a Markov Process},
	journal = {Bull. Math. Biol.},
	volume  = {79},
	number  = {12},
	pages   = {2905--2928},   
	year    = {2017},
	doi     = {10.1007/s11538-017-0356-4}
}

@article{Chao2019,
  title = {Evidence that the human cell cycle is a series of uncoupled,  memoryless phases},
  volume = {15},
  ISSN = {1744-4292},
  url = {http://dx.doi.org/10.15252/msb.20188604},
  DOI = {10.15252/msb.20188604},
  number = {3},
  pages = {MSB188604},
  journal = {Mol. Syst. Biol.},
  publisher = {Springer Science and Business Media LLC},
  author = {Chao,  Hui Xiao and Fakhreddin,  Randy I and Shimerov,  Hristo K and Kedziora,  Katarzyna M and Kumar,  Rashmi J and Perez,  Joanna and Limas,  Juanita C and Grant,  Gavin D and Cook,  Jeanette Gowen and Gupta,  Gaorav P and Purvis,  Jeremy E},
  year = {2019}
}

@article{Ham2024,
	author  = {Ham, Lucy and Coomer, Megan A. and {\"O}cal, Kaan and Grima, Ramon and Stumpf, Michael P. H.},
	title   = {A stochastic vs deterministic perspective on the timing of cellular events},
	journal = {Nat. Commun.},
	volume  = {15},
	number  = {1},
	pages   = {5286},
	year    = {2024},
	doi     = {10.1038/s41467-024-49624-z}
}

@book{vanKampen2007,
	author    = {van Kampen, N. G.},
	title     = {Stochastic Processes in Physics and Chemistry},
	edition   = {3rd},
	publisher = {Elsevier},
	address   = {Amsterdam},
	year      = {2007}
}

@article{CooperHelmstetter1968,
  author  = {Cooper, Stephen and Helmstetter, Charles E.},
  title   = {Chromosome replication and the division cycle of
             {\textit{Escherichia coli}} {B/r}},
  journal = {J. Mol. Biol.},
  volume  = {31},
  pages   = {519},
  year    = {1968}
}

@article{Bryant2014,
  author  = {Bryant, Jack A. and Sellars, Laura E. and Busby, Stephen J. W.
             and Lee, David J.},
  title   = {Chromosome position effects on gene expression in
             {\textit{Escherichia coli}} {K}-12},
  journal = {Nucleic Acids Res.},
  volume  = {42},
  pages   = {11383},
  year    = {2014}
}

@article{sousa1997,
  title={Modulation of gene expression through chromosomal positioning in Escherichia coli},
  author={Sousa, Carolina and {de Lorenzo}, V{\'\i}ctor and Cebolla, Angel},
  journal={Microbiology},
  volume={143},
  number={6},
  pages={2071--2078},
  year={1997},
  publisher={Microbiology Society}
}

@article{Swain2002,
  title = {Intrinsic and extrinsic contributions to stochasticity in gene expression},
  volume = {99},
  ISSN = {1091-6490},
  url = {http://dx.doi.org/10.1073/pnas.162041399},
  DOI = {10.1073/pnas.162041399},
  number = {20},
  journal = {Proc. Natl. Acad. Sci. U.S.A.},
  publisher = {National Academy of Sciences},
  author = {Swain,  Peter S. and Elowitz,  Michael B. and Siggia,  Eric D.},
  year = {2002},
  pages = {12795–12800}
}

@article{Elowitz2002,
  title = {Stochastic Gene Expression in a Single Cell},
  volume = {297},
  ISSN = {1095-9203},
  url = {http://dx.doi.org/10.1126/science.1070919},
  DOI = {10.1126/science.1070919},
  number = {5584},
  journal = {Science},
  publisher = {American Association for the Advancement of Science (AAAS)},
  author = {Elowitz,  Michael B. and Levine,  Arnold J. and Siggia,  Eric D. and Swain,  Peter S.},
  year = {2002},
  month = Aug,
  pages = {1183–1186}
}

@article{verma2026,
  title={Distributions of threshold crossing times of messenger RNA},
  author={Verma, Arunendra Kumar and Barman, Hillol Kumar and Rijal, Krishna and Das, Dibyendu},
  journal={bioRxiv},
  doi = {10.64898/2026.08.20.745891},
  year={2026},
  publisher={Cold Spring Harbor Laboratory}
}

@article{Burdett1996,
  title = {Truncated means and variances},
  volume = {52},
  number = {3},
  journal = {Econ. Lett.},
  publisher = {Elsevier BV},
  author = {Burdett,  Kenneth},
  year = {1996},
  pages = {263–267}
}

@article{Dileep2018,
  title = {Single-cell replication profiling to measure stochastic variation in mammalian replication timing},
  volume = {9},
  ISSN = {2041-1723},
  url = {http://dx.doi.org/10.1038/s41467-017-02800-w},
  DOI = {10.1038/s41467-017-02800-w},
  number = {1},
  pages = {428},
  journal = {Nat. Commun.},
  publisher = {Springer Science and Business Media LLC},
  author = {Dileep,  Vishnu and Gilbert,  David M.},
  year = {2018}
}

@misc{NIST:DLMF,
         key = "{\relax DLMF}",
       title = "{\it NIST Digital Library of Mathematical Functions}",
howpublished = "\url{https://dlmf.nist.gov/}, Release 1.2.8 of 2026-09-15",
         url = "https://dlmf.nist.gov/",
        year = "2026",
        note = "{F}.~W.~J. Olver, A.~B. {Olde Daalhuis}, D.~W. Lozier, B.~I. Schneider,
                R.~F. Boisvert, C.~W. Clark, B.~R. Miller, B.~V. Saunders,
                H.~S. Cohl, and M.~A. McClain, eds."
}

@book{Feller1966,
  author    = {Feller, William},
  title     = {An Introduction to Probability Theory and Its Applications},
  volume    = {2},
  edition   = {1},
  publisher = {John Wiley \& Sons},
  address   = {New York},
  year      = {1966}
}

@book{Feller1968,
  author    = {Feller, William},
  title     = {An Introduction to Probability Theory and Its Applications},
  volume    = {1},
  edition   = {3},
  publisher = {John Wiley \& Sons},
  address   = {New York},
  year      = {1968}
}

@book{CasellaBerger2002,
  author    = {Casella, George and Berger, Roger L.},
  title     = {Statistical Inference},
  edition   = {2},
  publisher = {Duxbury},
  address   = {Pacific Grove, CA},
  year      = {2002}
}

@article{Rijal2022,
  title = {Exact Distribution of Threshold Crossing Times for Protein Concentrations: Implication for Biological Timekeeping},
  volume = {128},
  ISSN = {1079-7114},
  url = {http://dx.doi.org/10.1103/PhysRevLett.128.048101},
  DOI = {10.1103/physrevlett.128.048101},
  number = {4},
  pages = {048101},
  journal = {Phys. Rev. Lett.},
  publisher = {American Physical Society (APS)},
  author = {Rijal,  Krishna and Prasad,  Ashok and Singh,  Abhyudai and Das,  Dibyendu},
  year = {2022}
}

@article{Jia2023,
  title = {Coupling gene expression dynamics to cell size dynamics and cell cycle events: Exact and approximate solutions of the extended telegraph model},
  volume = {26},
  ISSN = {2589-0042},
  url = {http://dx.doi.org/10.1016/j.isci.2022.105746},
  DOI = {10.1016/j.isci.2022.105746},
  number = {1},
  journal = {iScience},
  publisher = {Elsevier BV},
  author = {Jia,  Chen and Grima,  Ramon},
  year = {2023},
  month = Jan,
  pages = {105746}
}

@article{Dowling2014,
  title = {Stretched cell cycle model for proliferating lymphocytes},
  volume = {111},
  ISSN = {1091-6490},
  url = {http://dx.doi.org/10.1073/pnas.1322420111},
  DOI = {10.1073/pnas.1322420111},
  number = {17},
  journal = {Proc. Natl. Acad. Sci. U.S.A.},
  publisher = {National Academy of Sciences},
  author = {Dowling,  Mark R. and Kan,  Andrey and Heinzel,  Susanne and Zhou,  Jie H. S. and Marchingo,  Julia M. and Wellard,  Cameron J. and Markham,  John F. and Hodgkin,  Philip D.},
  year = {2014},
  month = Apr,
  pages = {6377–6382}
}

@article{Sukys2025,
  title = {Cell-cycle dependence of bursty gene expression: insights from fitting mechanistic models to single-cell RNA-seq data},
  volume = {53},
  ISSN = {1362-4962},
  url = {http://dx.doi.org/10.1093/nar/gkaf295},
  DOI = {10.1093/nar/gkaf295},
  number = {7},
  pages = {gkaf295},
  journal = {Nucleic Acids Res.},
  publisher = {Oxford University Press (OUP)},
  author = {Sukys,  Augustinas and Grima,  Ramon},
  year = {2025},
  month = Apr 
}

@article{Skinner2016,
  title = {Single-cell analysis of transcription kinetics across the cell cycle},
  volume = {5},
  pages = {e12175},
  ISSN = {2050-084X},
  url = {http://dx.doi.org/10.7554/eLife.12175},
  DOI = {10.7554/elife.12175},
  journal = {eLife},
  publisher = {eLife Sciences Publications,  Ltd},
  author = {Skinner,  Samuel O and Xu,  Heng and Nagarkar-Jaiswal,  Sonal and Freire,  Pablo R and Zwaka,  Thomas P and Golding,  Ido},
  year = {2016}
}

@article{Kalita2025,
  title = {An Hfq-dependent post-transcriptional mechanism fine tunes RecB expression in Escherichia coli},
  volume = {13},
  pages = {RP94918},
  ISSN = {2050-084X},
  url = {http://dx.doi.org/10.7554/eLife.94918.3},
  DOI = {10.7554/elife.94918.3},
  journal = {eLife},
  publisher = {eLife Sciences Publications,  Ltd},
  author = {Kalita,  Irina and Iosub,  Ira Alexandra and McLaren,  Lorna and Goossens,  Louise and Granneman,  Sander and El Karoui,  Meriem},
  year = {2025},
  month = Aug 
}

@article{Voichek2016,
  title = {Expression homeostasis during DNA replication},
  volume = {351},
  ISSN = {1095-9203},
  url = {http://dx.doi.org/10.1126/science.aad1162},
  DOI = {10.1126/science.aad1162},
  number = {6277},
  journal = {Science},
  publisher = {American Association for the Advancement of Science (AAAS)},
  author = {Voichek,  Yoav and Bar-Ziv,  Raz and Barkai,  Naama},
  year = {2016},
  month = Mar,
  pages = {1087–1090}
}

@article{Schwanhusser2011,
  title = {Global quantification of mammalian gene expression control},
  volume = {473},
  ISSN = {1476-4687},
  url = {http://dx.doi.org/10.1038/nature10098},
  DOI = {10.1038/nature10098},
  number = {7347},
  journal = {Nature},
  publisher = {Springer Science and Business Media LLC},
  author = {Schwanh\"{a}usser,  Bj\"{o}rn and Busse,  Dorothea and Li,  Na and Dittmar,  Gunnar and Schuchhardt,  Johannes and Wolf,  Jana and Chen,  Wei and Selbach,  Matthias},
  year = {2011},
  month = May,
  pages = {337–342}
}

@article{Yu2006,
  title = {Probing Gene Expression in Live Cells,  One Protein Molecule at a Time},
  volume = {311},
  ISSN = {1095-9203},
  url = {http://dx.doi.org/10.1126/science.1119623},
  DOI = {10.1126/science.1119623},
  number = {5767},
  journal = {Science},
  publisher = {American Association for the Advancement of Science (AAAS)},
  author = {Yu,  Ji and Xiao,  Jie and Ren,  Xiaojia and Lao,  Kaiqin and Xie,  X. Sunney},
  year = {2006},
  month = Mar,
  pages = {1600–1603}
}

@article{Loman2023,
  title = {Catalyst: Fast and flexible modeling of reaction networks},
  volume = {19},
  number = {10},
  journal = {PLoS Comput. Biol.},
  publisher = {Public Library of Science (PLoS)},
  author = {Loman,  Torkel E. and Ma,  Yingbo and Ilin,  Vasily and Gowda,  Shashi and Korsbo,  Niklas and Yewale,  Nikhil and Rackauckas,  Chris and Isaacson,  Samuel A.},
  editor = {Ouzounis,  Christos A.},
  year = {2023},
  month = Oct,
  pages = {e1011530}
}

@article{Mosteller1980,
  title = {Metabolism of individual proteins in exponentially growing Escherichia coli.},
  volume = {255},
  ISSN = {0021-9258},
  url = {http://dx.doi.org/10.1016/S0021-9258(19)85924-0},
  DOI = {10.1016/s0021-9258(19)85924-0},
  number = {6},
  journal = {J. Biol. Chem.},
  publisher = {Elsevier BV},
  author = {Mosteller,  R.D. and Goldstein,  R.V. and Nishimoto,  K.R.},
  year = {1980},
  month = Mar,
  pages = {2524–2532}
}

@article{Gupta2024,
  title = {Global protein turnover quantification in Escherichia coli reveals cytoplasmic recycling under nitrogen limitation},
  volume = {15},
  ISSN = {2041-1723},
  url = {http://dx.doi.org/10.1038/s41467-024-49920-8},
  DOI = {10.1038/s41467-024-49920-8},
  number = {1},
  pages = {5890},
  journal = {Nat. Commun.},
  publisher = {Springer Science and Business Media LLC},
  author = {Gupta,  Meera and Johnson,  Alex N. T. and Cruz,  Edward R. and Costa,  Eli J. and Guest,  Randi L. and Li,  Sophia Hsin-Jung and Hart,  Elizabeth M. and Nguyen,  Thao and Stadlmeier,  Michael and Bratton,  Benjamin P. and Silhavy,  Thomas J. and Wingreen,  Ned S. and Gitai,  Zemer and W\"{u}hr,  Martin},
  year = {2024}
}

@article{Gillespie1976,
  title = {A general method for numerically simulating the stochastic time evolution of coupled chemical reactions},
  volume = {22},
  number = {4},
  journal = {J. Comput. Phys.},
  publisher = {Elsevier BV},
  author = {Gillespie,  Daniel T},
  year = {1976},
  month = Dec,
  pages = {403–434}
}

@article{Gillespie2007,
  title = {Stochastic Simulation of Chemical Kinetics},
  volume = {58},
  number = {1},
  journal = {Annu. Rev. Phys. Chem.},
  publisher = {Annual Reviews},
  author = {Gillespie,  Daniel T.},
  year = {2007},
  month = May,
  pages = {35–55}
}

@article{Shahrezaei2008,
  title = {Analytical distributions for stochastic gene expression},
  volume = {105},
  number = {45},
  journal = {Proc. Natl. Acad. Sci. U.S.A.},
  publisher = {National Academy of Sciences},
  author = {Shahrezaei,  Vahid and Swain,  Peter S.},
  year = {2008},
  month = Nov,
  pages = {17256–17261}
}

@article{Walker2016,
  title = {Generation and filtering of gene expression noise by the bacterial cell cycle},
  volume = {14},
  ISSN = {1741-7007},
  number = {1},
  pages = {11},
  journal = {BMC Biol.},
  publisher = {Springer Science and Business Media LLC},
  author = {Walker,  Noreen and Nghe,  Philippe and Tans,  Sander J.},
  year = {2016},
  month = Feb 
}

@article{Chen2016,
  title = {The Genomic Landscape of Position Effects on Protein Expression Level and Noise in Yeast},
  volume = {2},
  number = {5},
  journal = {Cell Syst.},
  publisher = {Elsevier BV},
  author = {Chen,  Xiaoshu and Zhang,  Jianzhi},
  year = {2016},
  month = May,
  pages = {347–354}
}

@book{Billingsley1995,
  author    = {Billingsley, Patrick},
  title     = {Probability and Measure},
  edition   = {3rd},
  publisher = {John Wiley \& Sons},
  address   = {New York},
  year      = {1995}
}

\end{document}